\documentclass[twocolumn,english,pre,superscriptaddress,showpacs,longbibliography]{revtex4-1}
\usepackage[colorlinks=true,urlcolor=blue,citecolor=blue,linkcolor=blue]{hyperref} 
\usepackage[T1]{fontenc}
\usepackage[latin9]{inputenc}
\usepackage{amssymb}
\usepackage{graphicx}
\usepackage[dvipsnames]{xcolor}
\usepackage{amsmath}
\usepackage{mathrsfs}
\usepackage{float}
\usepackage{indentfirst}
\usepackage{txfonts}
\usepackage{babel}
\usepackage{siunitx}
\usepackage{tabularx}
\usepackage{booktabs} \newcommand{\ra}[1]{\renewcommand{\arraystretch}{#1}}

\makeatletter

\def\journal #1, #2, #3, 1#4#5#6{{\sl #1~}{\bf #2}, #3 (1#4#5#6) }

\def\eqa{\begin{eqnarray}}
\def\eea{\end{eqnarray}}
\newcommand{\eq}{\begin{equation}}
\newcommand{\ee}{\end{equation}}

\newcommand{\Eq}[1]{Eq.~(\ref{#1})}

\newcommand{\vphysical}{\boldsymbol{x}}
\newcommand{\vlatent}{\boldsymbol{z}}
\newcommand{\vlambda}{\boldsymbol{\lambda}}
\newcommand{\vp}{\boldsymbol{p}}
\newcommand{\vq}{\boldsymbol{q}}
\newcommand{\vP}{\boldsymbol{P}}
\newcommand{\vQ}{\boldsymbol{Q}}
\newcommand{\MA}{Monge-Amp\`ere }

\newcommand{\commentOut}[1]{}

\makeatother

\begin{document}

%\title{Neural Canonical Transformations}
\title{Neural Canonical Transformation with Symplectic Flows}
%\title{Neural Canonical Transformation for many-particle dynamical systems}
\author{Shuo-Hui Li}
%\email{contact_lish@iphy.ac.cn}
\affiliation{Institute of Physics, Chinese Academy of Sciences, Beijing 100190, China}
\affiliation{University of Chinese Academy of Sciences, Beijing 100049, China}
\author{Chen-Xiao Dong}
\affiliation{Institute of Physics, Chinese Academy of Sciences, Beijing 100190, China}
\affiliation{University of Chinese Academy of Sciences, Beijing 100049, China}
\author{Linfeng Zhang}
\email{linfengz@princeton.edu}
\affiliation{Program in Applied and Computational Mathematics, Princeton University, Princeton, NJ 08544, USA}
\author{Lei Wang}
\email{wanglei@iphy.ac.cn}
\affiliation{Institute of Physics, Chinese Academy of Sciences, Beijing 100190, China}
%\affiliation{CAS Center for Excellence in Topological Quantum Computation, University of Chinese Academy of Sciences, Beijing 100190, China}
\affiliation{Songshan Lake Materials Laboratory, Dongguan, Guangdong 523808, China}

\begin{abstract}
Canonical transformation plays a fundamental role in simplifying and solving classical Hamiltonian systems. 
Intriguingly, it has a natural correspondence to normalizing flows with a symplectic constraint. 
Building on this key insight, we design a neural canonical transformation approach to automatically identify independent slow collective variables in general physical systems and natural datasets. 
%a latent representation with an independent harmonic oscillator Hamiltonian. 
%Correspondingly, the phase space density of the physical system flows towards a factorized Gaussian distribution in the latent space.
%Thus, this latent representation has a structure defined by different frequency modes, with which we can extract patterns from.
%Since the canonical transformation preserves the Hamiltonian evolution, the model captures nonlinear collective modes in the learned latent representation.
We present an efficient implementation of symplectic neural coordinate transformations and two ways
%, variational free energy and density estimation, 
to train the model based either on the Hamiltonian function or phase space samples. 
The learned model maps physical variables onto an independent representation where collective modes with different frequencies are separated, which can be useful for various downstream tasks such as compression, prediction, control, and sampling. 
%The variational free energy calculation is based on the analytical form of physical Hamiltonian. 
%While the phase space density estimation only requires samples in the coordinate space for separable Hamiltonians. 
We demonstrate the ability of this method first by analyzing toy problems and then by applying it to real-world problems, such as identifying and interpolating slow collective modes of the alanine dipeptide molecule and MNIST images.
%Unsupervised identification of independent slow modes is useful for various downstream tasks such as sampling and compression. 
%including continuous symplectic flow, linear symplectic flow, and neural point transformations.
\end{abstract}
\maketitle

\section{Introduction}

%symplectic symmetry 
The inherent symplectic structure of classical Hamiltonian mechanics has profound theoretical and practical implications~\cite{Arnold1989}. %many-particle MD, importance and challenges 
For example, the symplectic symmetry underlies the Liouville's theorem~\cite{liouville1838note}, which states that the phase space density is incompressible under the Hamiltonian evolution. 
Canonical transformations which preserves the symplectic symmetry in the phase space have been a key technique for simplifying and solving Hamiltonian dynamics.
%Canonical transformations of phase space variables preserve the symplectic symmetry, it has been a key technique for simplifying and solving Hamiltonian dynamics. 
%In fact, both Hamiltonian evolution and canonical transformations are symplectic flows in the phase space. 
%can be used to identify the integrals of motion of the system. Moreover,
%One can solve or extract integral of motion of a Hamiltonian system by finding a suitable canonical transformation that simplifies the dynamics. 
%identify the integral motion
%MD and its importance and challenges
Respecting the intrinsic symplectic symmetry of Hamiltonian systems is also crucial for stable and energy-conserving numerical integration schemes~\cite{Feng2011} which play central roles in the investigations of celestial mechanics and molecular dynamics. 

%MD is a vital tool for understanding complex phenomena in physics, chemistry and biological systems, and for practical material drug design. For these purposes, one is particularly interested in slow collective modes emerged from microscopic degrees of freedom. However, a central difficulty of MD simulation is to bridge the huge separation between time scales of a many-particle system. For example, protein folding occurs at a time scale of $0.1$s. While the time step of MD simulation is limited femtoseconds to correctly capture the fast chemical bond vibration. A strategy to address the time scale problem is to identify a few slow collective variables and try to enhance sampling of these collective variables~\cite{Noe}. 

Molecular dynamics (MD) simulation investigates the dynamical and statistical properties of matter by integrating the equations of motion of a large number of atoms. 
MD is a vital tool for understanding complex physical, chemical, and biological phenomena, as well as for practical applications in material discovery and drug design. 
Modern MD simulation generates huge datasets, which encapsulate the full microscopic details of the molecular system~\cite{Lindorff-Larsen2010}. 
However, this also poses challenges to the development of data analysis tools. 
In particular, one typically interests in the emerging slow modes, which are often related to the collective property of the system. 
Moreover, identifying such degrees of freedom is also crucial for enhanced sampling of molecular conformations. See Refs.~\cite{Noe, Wang2019} for recent reviews. 

Machine learning techniques provide promising solutions to these problems in MD. 
For example, the time-lagged independent component analysis (t-ICA)~\cite{Molgedey1994, Klaus-Robert97, Klaus-Robert98, Perez-Hernandez2013, Schwantes2013} separates a linear mixture of independent time-series signals. 
The approach shows a close connection to the dynamic mode decomposition scheme developed in the fluid mechanics community~\cite{Schmid2010,Klus2018}. Many of these linear analysis methods have nonlinear generalizations based on kernel approaches~\cite{Klaus-Robert98Nonl, Klaus-Robert03}. More recently, several deep learning approaches have also been proposed to identify nonlinear coordinate transformation of dynamical systems~\cite{Mardt2018, Wehmeyer2018, Hernandez2018a, Sultan, Lusch2018}. Parallel to these efforts, it is also an active research direction to extract slow features in general time-series data~\cite{Klaus-Robert98Nonl, Wiskott2002a, Pfau2018a} within the machine learning community. 

%canonical transformation + DL
In this paper, we develop a different approach by exploiting the inherent connection between canonical transformation and normalizing flows~\cite{Kobyzev2019, Papamakarios2019}. 
%Having realized that canonical transformations are phase space flow models with a symplectic condition, 
We design a class of learnable neural canonical transformations to simplify complex Hamiltonian dynamics of the physical variables towards independent collective motions in the transformed phase space. 
Correspondingly, the canonical transformation also reduces complex phase space densities towards an independent Gaussian prior distribution. After learning, one can directly control nonlinear collective variables with different frequencies by tuning independent collective variables in the latent space of the normalizing flows. 
We present learning algorithms and discuss applications of the neural canonical transformation on the extraction of slow collective variables of the physical and realistic dataset. 
We stress that most techniques targeting at extracting dynamical information require time-series data. 
However, in the present approach, the dynamical information is imposed on the structure of the neural canonical transformation, so that the training scheme does not necessarily follow a specific time step, and the data sample may come from other types of sampling methods, such as Monte Carlo, biased dynamics, etc.

There have been related research works exploiting the symplectic property in machine learning tasks. Ref.~\cite{Mattheakis} solves Hamiltonian equations using neural networks. Refs.~\cite{Greydanus2019} learns a Hamiltonian dynamics from observed data using neural networks. 
Both studies found that exploiting the symplectic structure in the learning helps in boosting the performance. More recently, there have been more preprints on related topics~\cite{Sanchez-Gonzalez2019,Zhong2019,Chen2019,Rezende2019,Toth2019} which also aim at improving the performance in machine learning tasks by imposing physics motivated inductive biases in the neural network design. Our work finds the closest connection to Ref.~\cite{Bondesan2019}, which investigates classical integrable systems using symplectic neural networks. Our paper targets at more general settings and aims to identify nonlinear slow collective modes of complex systems. In addition, we also note that there are efforts on learning neural networks for the force fields of molecular dynamics~\cite{behler2007generalized,schutt2017quantum,han2017deep,zhang2018deep,zhang2018end,Chmiela2018,Chmielae1603015}, wherein imposing the physical invariance is also crucial. 

The organization of the paper is as follows. In section~\ref{sec:theory}, we reveal the key connection of canonical transformation to normalizing flows with the symplectic condition. In section~\ref{sec:nct}, we present the design and training of the symplectic neural networks for canonical transformation. We also discuss potential applications of the neural canonical transformation. In section~\ref{sec:examples} we demonstrate applications of the neural canonical transformation to toy problems and realistic data. Finally, we discuss possible prospects of neural canonical transformation in Sec.~\ref{sec:discussions}. 

\section{Theoretical background} \label{sec:theory}
We review the canonical transformation and its connection to the normalizing flow model. %This section sets the notations for later discussions. 

\subsection{Canonical transformation of Hamiltonian systems}
We denote the canonical variables, namely the momenta and coordinates of a Hamiltonian system, as a row vector with $2n$ elements $\vphysical \equiv  (\vp, \vq)$. The Hamiltonian equation can be concisely written as $ \dot{\vphysical}  = \nabla_{\vphysical} H(\vphysical) J$, where $\dot{\vphysical}$ denotes time derivative of the canonical variables. $H(\vphysical)$ is the Hamiltonian function and $J = \left(\begin{array}{cc} & I \\ -I & \end{array}\right)$ is a $2n\times 2n$ symplectic metric matrix. 

The canonical transformation is a bijective mapping from the original canonical variables to a new set of canonical variables, i.e. $\mathcal{T}: \vphysical \mapsto \vlatent \equiv  (\vP, \vQ) $, whose Jacobian matrix $M_{ij} = \nabla_{x_j} z_i$
%\frac{\partial \zeta_i}{\partial \eta_j}
satisfies the symplectic condition 
\begin{equation}
M J M^T = J. 
\label{eq:symplecticcondition}
\end{equation}
%We denote by $\vphysical$ the original variables in the physical phase space, with the Hamiltonian $H$, and by $\vlatent$ the transformed, usually simplified, phase space variables, with the Hamiltonian $K$, respectively.
%This canonical transformation essentially establishes a bijective mapping between the original Hamiltonian H and the transformed Hamiltonian K. Thus, according to Hamiltonian dynamics, also establishes a bijective mapping between the time evolution trajectory in original phase space and transformed phase space, as illustrated in Fig.~\ref{fig:concept}

Canonical transformation preserves the Hamiltonian equation, i.e. one has $\dot{\vlatent} = \nabla_{\vlatent} K(\vlatent) J$, where $K(\vlatent) = H \circ \mathcal{T}^{-1}(\vlatent)$ is a transformed Hamiltonian in terms of the new phase space variables. 
The canonical transformation establishes a bijective mapping between the Hamiltonian trajectories in the original and the transformed phase spaces. Thus, one can search for canonical transformations which simplify and even solve the Hamiltonian dynamics. One of such searching strategy is to compose elementary canonical transformations since the symplectic condition \Eq{eq:symplecticcondition} form a group. 
%, and the inverse of a canonical transformation is also a canonical transformation. %, i.e., one has $\mathcal{T} \left ( \vphysical + \int \dot{\vphysical}\, dt \right ) =  \vlatent + \int \dot{\vlatent}\, dt  $ for an arbitrary evolution time. 
%Therefore, although the equations of motion may appear different in the physical and the latent spaces, they describe the same Hamiltonian evolution as illustrated in Fig.~\ref{fig:concept}. 

\begin{figure}
    \begin{center}
    \includegraphics[width=\columnwidth, clip]{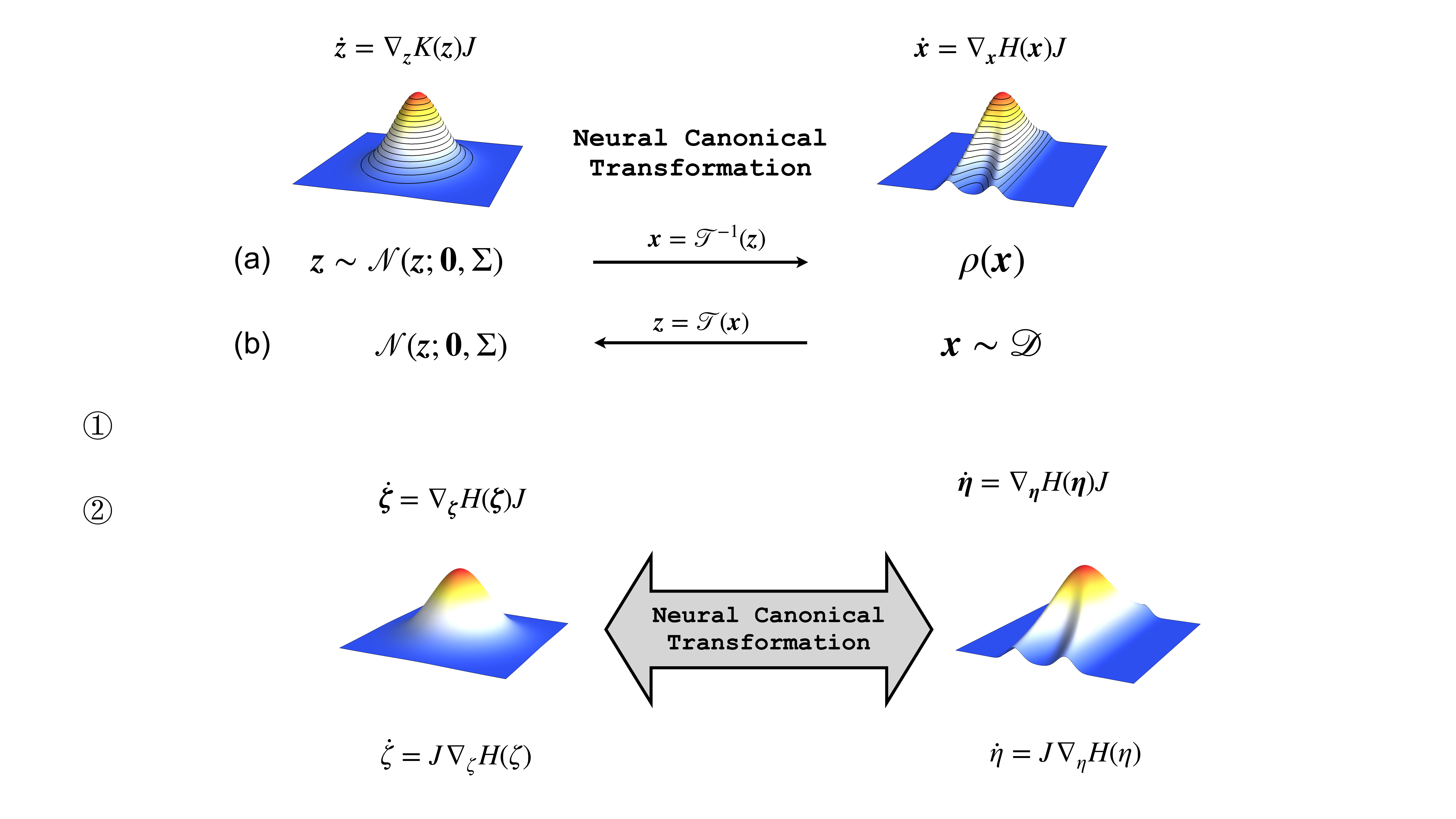}
    \end{center}
    \caption{A neural canonical transformation maps between the latent variables $\vlatent$ and physical variables $\vphysical$ via a symplectic neural network. The transformation preserves the Hamiltonian equation and connects the phase space trajectories in the physical and the latent spaces. %Moreover, the transformation deforms the phase space density in a volume-preserving manner. 
%We learn the symplectic transformation together with the harmonic frequencies in \Eq{eq:priorhamiltonian} of the latent space Hamiltonian. 
There are two ways to train the neural canonical transformation: (a) variational free energy based on the Hamiltonian [\Eq{eq:pddloss}]. (b) density of phase space estimation based on data [\Eq{eq:mleloss}].}
 \label{fig:concept}
\end{figure}

%In general, a dynamical system explores the phase space ergodically in the long time limit. 
%Ergodicity seems to be an assumption for most cases. See https://en.wikipedia.org/wiki/Ergodic_hypothesis. So I rephrased this sentence.
%\subsection{Flow-based generative models}
\subsection{Normalizing flow models}

Generative modeling aims at modeling the joint distribution of complex high dimensional data~\cite{Goodfellow2016a}. 
A generative model can capture the key variations of the dataset and draw samples efficiently from the learned probability distribution. 
A large class of generative models achieves these goals by learning a transformation from a simple latent distribution to a complex physical distribution. %In most cases, this kind of transformation is irreversible.
%that encodes collective patterns of the data. 
%For example, the generative adversarial networks (GAN) transform latent variables which follow a simple prior distribution to the target distribution~\cite{Goodfellow2014a}; the variational autoencoders (VAE) model the conditional probabilities between the target distribution and the latent distribution using an encoder and a decoder network~\cite{Kingma2013}.
Examples well-known to the machine learning community include the generative adversarial networks (GAN)~\cite{Goodfellow2014a}, the variational autoencoders (VAE) model~\cite{Kingma2013} and the normalizing flow models~\cite{Kobyzev2019, Papamakarios2019}. 

The normalizing flow models, or the flow-based generative models, are particularly suitable for our purpose since they employ a bijective mapping from the latent space $\vlatent$ to the target space $\vphysical$ for the probability transformation. Essentially, the normalizing flow models perform change of variables to induce transformations in the probability densities. Typically, the normalizing flow model transforms a simple prior distribution of the latent variables following, e.g., the Gaussian distribution $\mathcal{N}(\vlatent; \boldsymbol{0}, \Sigma)$ with zero mean and covariance $\Sigma$ to the more complex distribution of the realistic data, and vice versa. %, hence, also has the name "normalizing" flow.  

%\comment{Also, the change of probability in the transformation is exactly solvable via some limitations put on the change of variables formula. Such limitations often may  harm the learning abilities of the models. Thus, one critically task of designing one such model is the balance of tractability and scalability.}
There have been large flavors of normalizing flow models~\cite{NormalizingFlow, Dinh2014, Dinh2016a, Kingma2018, Grathwohl2018} that achieved nice performance in machine learning applications. 
Compared to GAN and VAEs, the normalizing flow models %can be seen as a combined generative model of both~\cite{TianHan2018}, thus has the merit of both and several other
have appealing features such as tractable likelihood for any given data $\vphysical$ and exact reversibility between $\vlatent$ from $\vphysical$. These features are particularly attractive for principled and quantitative scientific applications such as learning renormalization group flow~\cite{Li2018z}, holographic mapping~\cite{Hu2019}, Monte Carlo sampling~\cite{Song2017, Levy2017, Albergo2019}, molecular simulations~\cite{BoltzmannGenerator}, and spin glasses~\cite{2001.00585}. Interestingly, in the continuous-time limit, the normalizing flow model exhibits intriguing connections to a variety of topics including dynamical systems, ordinary differentiation equations, optimal transport theory and fluid dynamics~\cite{Weinan2017, NeuralODE, Zhang2018r}.

\subsection{Connections between canonical transformation and normalizing flow  models} \label{sec:connection}

%Based on the ergodicity hypothesis and the Liouville's theorem, the phase space points along the evolution trajectory, namely momenta and coordinates, can be considered as samples in the phase space. One can access the ensemble average of physical properties at equilibrium by simply computing the time average along the trajectory.  
Since canonical transformations are changes of variables in the phase space, they can be naturally parameterized and learned as normalizing flow models with the added symplectic condition. 
Moreover, it is instructive to consider the probabilistic interpretation of this change-of-variables. %canonical transformations. %have a besides its usual dynamical consequences. 
Consider the phase space density in the canonical ensemble $\pi(\vphysical) = e^{-\beta H(\vphysical)}/Z$, where $Z = \int d \vphysical e^{-\beta H(\vphysical)}$ is the partition function, $\beta = k_B T$ is the inverse temperature. Change-of-variables to a simplified Hamiltonian function $K(\vlatent)$ implies reaching a simplified density in the latent phase space $ e^{-\beta K(\vlatent)}/Z $. Thus, one can regard the canonical transformation as a flow-based generative model connecting the physical and latent phase spaces and the associated phase space densities. There is, however, a crucial symplectic constraint in \Eq{eq:symplecticcondition} on the transformation compared to ordinary normalizing flow models. From the generative modeling perspective, the additional symplectic condition further restricts the expressibility the network. On the other hand, the symplectic inductive bias offers physical guarantee and interpretability on the training results, e.g., the network will always be a canonical transformation that preserves the dynamics in the latent space. Table~\ref{tab:connection} summarizes the connection between canonical transformation and normalizing flow. 

This key insight has several profound consequences. First, one can search for canonical transformations by learning the flow models in the phase space, which simplifies complex Hamiltonians both in the statistical and in the dynamical senses. Second, the learned latent spaces attain the physical meaning of transformed canonical variables which can be useful for various downstream tasks. Lastly, the symplectic neural network necessarily has the volume-preserving property, which can be computationally efficient since the Jacobian determinant is always unity by construction. 

\begin{table}\centering
\ra{1.4}
    \scalebox{1.1}{
        \begin{tabularx}{0.9\columnwidth}{X X}\toprule
 \textbf{Canonical Transformation} & \textbf{Normalizing Flow} \\  \hline
 $\vphysical=(\vp, \vq)$   & physical variables \\
 $\vlatent=(\vP, \vQ)$   & latent  variables \\
 $\vphysical \leftrightarrow \vlatent $ &  symplectic flow \\
 $e^{-H(\vphysical)}/Z$   & physical phase space density  \\
  $e^{-K(\vlatent)}/Z$   & latent phase space density \\
\bottomrule
        \end{tabularx}
    }
\caption{The correspondence of a canonical transformation and normalizing flow. See Sec.~\ref{sec:connection} for explanations of the connection. \label{tab:connection}}
\end{table}

\section{Canonical transformation using normalizing flow  models} \label{sec:nct}

It is usually difficult to devise useful canonical transformations for generic Hamiltonians since it typically involves solving a large set of coupled nonlinear equations.
%analytically is usually not manageable for questions of practical interests. 
However, building on the connections of canonical transformation and normalizing flow models~\cite{Kobyzev2019, Papamakarios2019}, we can construct a family of expressive canonical transformations with symplectic neural networks and train them with optimization techniques. 

To train the model, one can follow either the variational approach or the data-driven approach. As a result, the neural canonical transformation helps simplify the dynamics and identify nonlinear slow collective variables of complex Hamiltonians. %We show that using reversible generative networks such as the flows~\cite{NormalizingFlow, Dinh2014, Dinh2016a, Kingma2018} one can construct highly efficient and flexible neural canonical transformations. 
%One does not need ot solve linear system for these transformations. 

\subsection{Model Architectures}

As a flow-based generative model, the neural canonical transformation consists of a symplectic network and a prior distribution which corresponds to the transformation and the target phase density distribution respectively. %We use neural networks to parametrize learnable canonical transformations which satisfy \Eq{eq:symplecticcondition}. 
%we pay special attention that these are scalable to large dataset and dimensions since a general class of the canonical transformations induced by a generation function may needs solve linear equations involving the Jacobian, which is then not so scalable. 
%\subsubsection{Neural Symplectic Transformations}
In the most general setting, the canonical transformation can even mix the momenta and coordinates. Here, we restrict ourselves to point transformations~\cite{Sussman} for balanced flexibility and interpretability. We list several other possible implementations of the neural canonical transformations in Appendix~\ref{sec:flows}. We note that one can compose symplectic neural networks to form more expressive canonical transformations.  

\subsubsection{Neural point transformations} \label{sec:neuralpointtransformation}

%In classical mechanics, canonical transformation usually takes the form of generator functions. Here, we use one kind of such generator function, point transformation, to parameterize a flow-based model. 
In the point transformation, one performs a nonlinear transformation to the coordinates $\vq$ and a linear transformation to the momenta $\vp$ accordingly,
\begin{eqnarray}
\vQ &=& \mathcal{F}(\vq) \label{eq:coord}, \\
\vP &=& \vp \left(\nabla_{\vq} \vQ  \right)^{-1} = \vp \nabla_{\vQ} \vq .   \label{eq:momentum} 
%\frac{\partial q}{\partial Q}.
\end{eqnarray}
The overall transformation in the phase space $\mathcal{T}: \vphysical=(\vp, \vq)  \mapsto \vlatent=(\vP, \vQ)  $ satisfies the symplectic condition \Eq{eq:symplecticcondition}~\cite{Sussman}. Training of point transformation will involve both momenta and coordinates in the phase space. 
Since the coordinate transformation \Eq{eq:coord} is independent of momenta, we can use the resulting coordinates $\vQ$ alone as a set of collective coordinates.
%In the second equality of \Eq{eq:momentum} we have exploited the reversibility of the coordinate transformation. Thus, this transformation can be turned into a flow-based neural network. Interestingly, the momentum transformation has the form of a vector-Jacobian product, which is commonly implemented for the reverse mode automatic differentiation~\cite{Baydin2018}. This is intuitively understandable since the momentum is covariant under the coordinate transformation. 

The coordinate transformation $\mathcal{F}:\vq \mapsto \vQ$ in \Eq{eq:coord} can be any nonlinear bijective mapping. We implement it with a real-valued non-volume preserving (real NVP) network~\cite{Dinh2016a}, which is a typical normalizing flow model~\cite{Kobyzev2019, Papamakarios2019}. The momenta transformation has the form of a vector-Jacobian product, which is commonly implemented for the reverse mode automatic differentiation~\cite{Baydin2018}. So we can leverage the automatic differentiation mechanism for the momentum transformation. % $\vP = \nabla_{\vQ} \left(\vp \cdot\mathcal{F}^{-1}(\vQ)\right)$. 
In practice, we run the coordinate transformation first forwardly and then reversely for $\vq=\mathcal{F}^{-1}(\vQ)$. Then, we compute its inner product with the initial momenta $\vp \cdot\vq$. Finally, we compute the derivative of the scalar with respect to $\vQ$ to obtain the transformed momenta $\vP$ in \Eq{eq:momentum}.

%One can similarly define point transformation for the momenta. However we find for practice application the coordinate transformations are more relevant.
%In the second equality of \Eq{eq:momentum} we have exploited the reversibility of the coordinate transformation. Thus, this transformation can be turned into a flow-based neural network. Interestingly, the momentum transformation has the form of a vector-Jacobian product, which is commonly implemented for the reverse mode automatic differentiation~\cite{Baydin2018}. This is intuitively understandable since the momentum is covariant under the coordinate transformation. 

\subsubsection{Latent space Hamiltonian and prior distribution}
%The neural canonical transformation Eqs~(\ref{eq:coord}, \ref{eq:momentum}) maps the physical variables to the latent space. 
We assume the transformed Hamiltonian in the latent space has the simple form of an independent harmonic oscillator
 \begin{equation}
K(\vlatent) = \sum_{k=1}^n \frac{ P_k^2 + \omega_k^2 Q_k^2}{2},%= -\frac{\vlatent \Sigma^{-1} \vlatent^T}{2} 
\label{eq:priorhamiltonian}
\end{equation}
where $\omega_k$ are learnable frequencies for each pair of conjugated canonical variables. Without loss of generality, we set the inverse temperature in the latent space to be one.
Therefore, in terms of canonical density distribution, the prior density in the latent space is an independent Gaussian $\mathcal{N}(\vlatent ; \boldsymbol{0}, \Sigma) =  \frac{\prod_{k=1}^n \omega_k}{(2\pi)^n}  e^{-K{(\vlatent)}}$, where $\Sigma = \mathrm{diag}(\underbrace{1, \ldots, 1}_{n}, \underbrace{\omega_1^{-2}, \ldots, \omega_n^{-2}}_{n})$ is a diagonal covariance matrix. In this setting, each pair of canonical variables in the latent space corresponds to an independent collective mode with a learnable frequency. Thus, after training one can select a desired number of slow modes according to the frequencies. %Then, the learned $\omega_k$ gives the frequency of this very pattern/mode. 

\subsection{Training Approaches}
The principle for training is to match the phase space density of the generative model $\rho$ to the target density $\pi$. Depending on specific applications one may either have direct access to the Hamiltonian function or the samples from the target distribution. Thus, we devised two training schemes of the neural canonical transformation based on variational calculation and the data-driven approach respectively.
%We devise suitable objective functions in below. 

\subsubsection{Variational approach} \label{eq:variational}
We can learn the canonical transformation based on the analytical expression of the physical Hamiltonian. 
For this purpose, we minimize the variational free energy 
\begin{equation}
%\mathcal{L}_\mathrm{PDD} = \mathbb{KL}\left( q \Vert \frac{e^{-H(p, q)/T}}{Z} \right)
\mathcal{L} = \int d \vphysical \, \rho(\vphysical) \left[ \ln \rho (\vphysical) + \beta H(\vphysical ) \right].
\label{eq:pddloss}
\end{equation}
This objective function is upper bounded by the free energy since $\mathcal{L} + \ln Z = \mathbb{KL}\left( \rho \Vert \pi \right)\ge 0$, where the Kullback-Leibler (KL) divergence is a non-negative measure of the dissimilarity between the model and the target distributions. The equality is reached only when two distributions match each other. 
The objective function of this form was recently employed in the probability density distillation of generative models~\cite{Oord2017}. 

To evaluate \Eq{eq:pddloss} we first draw samples from the normal distribution and then scale them according to the frequencies in the prior distribution to obtain $\vlatent \sim \mathcal{N}(\vlatent; \boldsymbol{0}, \Sigma)$. 
Next, we pass the samples through the symplectic transformation to obtain $\vphysical= \mathcal{T}^{-1}(\vlatent)$ as shown in Fig.~\ref{fig:concept}(a). Since the symplectic transformation is volume-preserving, the probability density of the produced samples read $\rho (\vphysical ) = \mathcal{N}(\vlatent; \boldsymbol{0}, \Sigma)$. The objective function is estimated on these samples as $\mathcal{L} = \mathbb{E}_{\vphysical\sim \rho(\vphysical)}[ \ln \rho(\vphysical) + \beta H(\vphysical)]$. To minimize the objective function we compute gradient over such sampling procedure with the reparameterization trick~\cite{Kingma2013}, which is an unbiased and a low variance gradient estimator for the learnable parameters~\cite{Mohamed2019}. 

\subsubsection{Maximum likelihood estimation}
Alternatively, one can also learn the canonical transformation in a purely data-driven approach. 
Assuming one already has access to independent and identically distributed samples from the target distribution $\pi(\vphysical)$, one can learn the neural canonical transformation with the maximum likelihood estimation on the data. 
This amounts to perform the density estimation in the phase space with a flow-based probabilistic generative model. The goal is to minimize the negative log-likelihood (NLL) on the dataset $\mathcal{D} = \{ \vphysical \}$
\begin{equation}
\mathrm {NLL} = -  \mathbb{E}_{\vphysical \sim \mathcal{D}} \left[ \ln \rho(\vphysical) \right],
\label{eq:mleloss}
\end{equation}
which reduces the observed phase space density and the model density $\mathbb{KL}\left( \pi \Vert \rho \right)$ based on empirical observations. 
To train the network we run the transformation from the physical to latent space as shown in Fig.~\ref{fig:concept}(b) and compute the model density $\rho (\vphysical) = \mathcal{N}(\vlatent=\mathcal{T}(\vphysical); \boldsymbol{0}, \Sigma)$. 

The density estimation \Eq{eq:mleloss} requires the phase space data, which involves both the coordinates and the momenta information. This appears to pose difficulties for applications to MD data which typically only contains the trajectory in the coordinate space. Fortunately, the momenta and coordinates distribution are factorized for separable Hamiltonians encounters in most MD simulations. That is, the momenta follow an independent Gaussian distribution whose variances depend on the atom masses and temperature. Therefore, one can exploit this fact and augment the training dataset by sampling momenta data directly from a Gaussian distribution. 
%Note that joint training for a factorized phase space distribution is meaningful since the symplectic condition relates the transformations for the momenta and coordinates. 

\subsection{Applications}

Neural canonical transformation learns a latent representation with independent modes and simplified dynamics. 
In principle, the learned representation is useful for the prediction, control, and sampling of the original system. 
We list a few concrete applications below. 

\subsubsection{Thermodynamics and excitation spectra}

Since the training approach of Sec.~\ref{eq:variational} satisfies the variational principle, the loss function \Eq{eq:mleloss} provides an upper bound to the physical free energy of the system. 
Besides, one can also estimate entropy and free energy differences of the Hamiltonian with different parameters. 
Similar variational free energy calculation of statistical mechanics problems using deep generative models have been carried out recently in Refs.~\cite{Li2018z, Zhang2018r, Wu2018f, Hu2019, Albergo2019, BoltzmannGenerator, 2001.00585}. In particular, Ref.~\cite{BoltzmannGenerator} has obtained encouraging results for sampling equilibrium molecular configurations. The present approach differs since it works in the phase space which involves both momenta and coordinates which allows extract dynamical information in addition to statistical properties.  

Since the neural canonical transformation preserves the Hamiltonian dynamics of the system, the learned frequencies in the latent space \Eq{eq:priorhamiltonian} reflect the intrinsic time scale of the target problem. 
In this way, the present approach captures coherent excitation of the system in the latent space harmonic motion. 
One may also estimate the spectral density based on the learned frequencies. 

%Although the density estimation calculation does not directly provide absolute value of the free energy, the trained generative models provide likelihood of each configuration. This can be used to estimate the 

\subsubsection{Identifying collective variables from slow modes}

Since the neural canonical transformation automatically separates dynamical modes with different frequencies in the latent space, %which are synchronized with the actual physical motion,
one can extract nonlinear slow  modes of the original physical system by selecting the latent variables with small frequencies. 

The neural canonical transformation differs fundamentally with these general time-series analysis approaches~\cite{Molgedey1994, Perez-Hernandez2013, Schwantes2013,Schmid2010, Klus2018} which do not exploit the domain-specific symplectic symmetry of the Hamiltonian systems. 
Another fundamental difference is that the canonical transformation is performed in the phase space which contains both momenta and coordinates, rather than for the time sequence of the coordinates. 
Since the explicit time information was never used in the present approach, there is no need to choose the time lag hyperparameter as in the t-ICA and related approaches. 
Lastly, variational training of the transformation also allows one to identify the canonical transformation directly from the microscopic Hamiltonian even without the time-series data.

We note that in practice, exact dynamical information is usually lost when one cares only about the structural, or static, information of a system.
Sophisticated thermosetting and enhanced sampling techniques are used to accelerate the sampling, but, meanwhile, the dynamics is destroyed.
In this case, MD plays the role of a sampler, rather than a real-time simulator, yet dynamical information can be extracted from statistical data of Hamiltonian systems with the neural canonical transformation approach.
 
\section{Examples}~\label{sec:examples}
We demonstrate the application of the neural canonical transformation with concrete examples. We start from simple toy problems and then move on to more challenging realistic problems. In all examples, the trainable parts of the network are the coordinate transformation $\mathcal{F}$ [\Eq{eq:coord}] and the latent space frequencies [\Eq{eq:priorhamiltonian}]. The code implementation is publicly available at~\cite{Github}. 

\subsection{Ringworld} \label{sec:ringworld}

\begin{figure}
    \begin{center}
    \includegraphics[width=\columnwidth, trim={0cm 0cm 0cm 1cm}, clip]{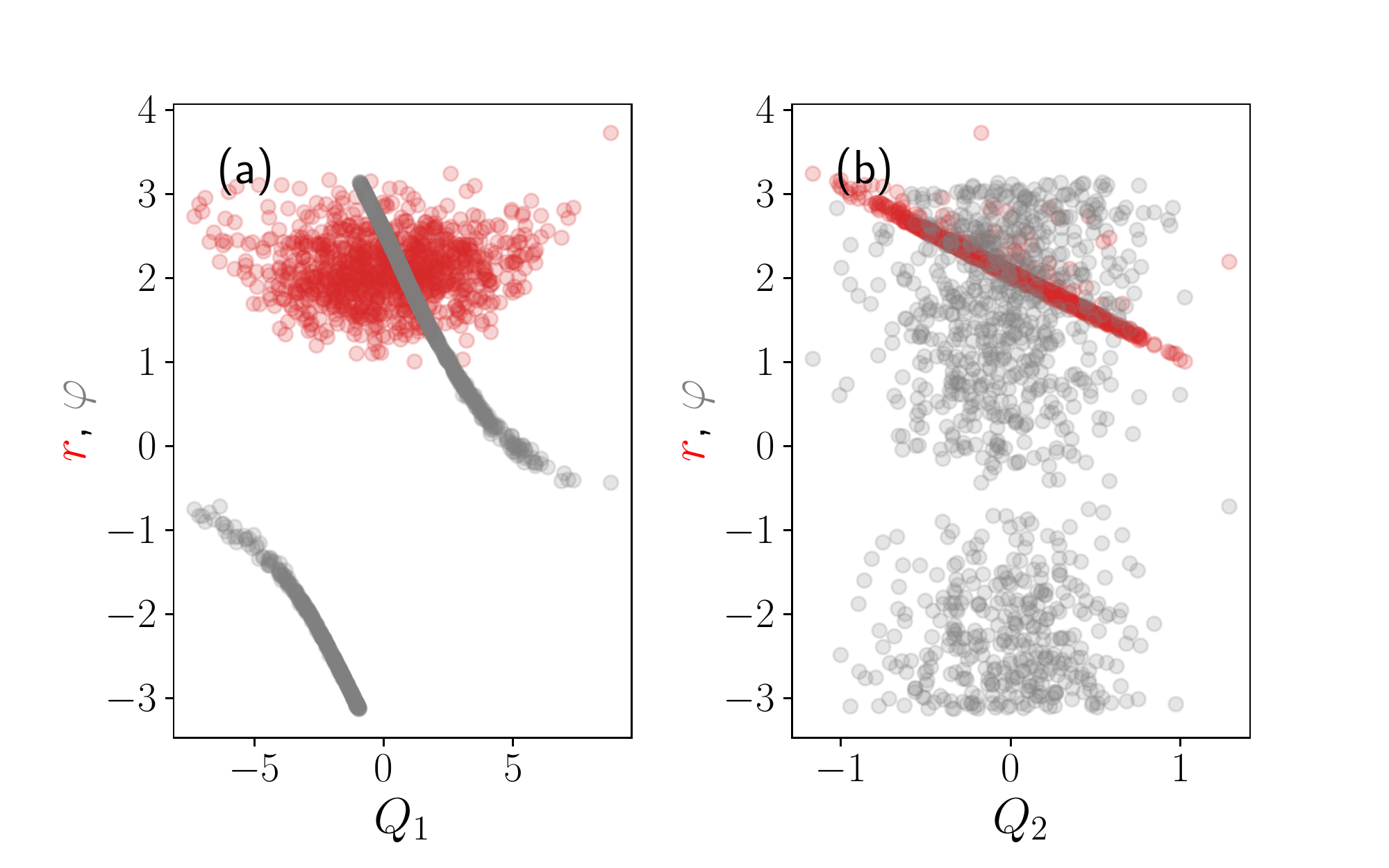} 
    \end{center}
    \caption{The Ringworld samples of Sec.~\ref{sec:ringworld}  projected to the plane of learned latent coordinates and the polar coordinates.}\label{fig:ring2d}
\end{figure}

First, we consider a two-dimensional toy problem with the Hamiltonian $H = \frac{1}{2}\left(p_1^2 + p_2^2 \right) + \left(\sqrt{q_1^2 + q_2^2}-2\right)^2/0.32$~\cite{Song2017}. 
Canonical distribution of this Hamiltonian resides in four-dimensional phase space. 
The canonical ensemble samples from $\pi(\vphysical)=e^{-H(\vphysical)}/Z$ are confined in a manifold embedded in the phase space due to the potential term. 
In the Euclidean space, the coordinates are correlated. 

Taking the Hamiltonian and training it with the variational approach, we obtain a neural point transformation from the original variables to a new set of canonical variables. 
Figure~\ref{fig:ring2d} shows the samples projected onto the plane of latent coordinates $Q_k$ and the polar coordinate variables $\varphi = \arctan(q_2/q_1),  r= \sqrt{q_1^2 + q_2^2}$. 
One observes a significant correlation between the slowest variable $Q_1$ and the polar angle $\varphi$, 
while the other transformed coordinate $Q_2$ shows a strong correlation with the radius variable $r$. 

This example demonstrates that, as a bottom line, the neural point transformation can automatically identify nonlinear transformation (such as polar coordinates) of the original coordinates. 
In the learned representation, the dynamics of each degree of freedom becomes independent. 

\subsection{Harmonic chain}\label{sec:harmonic}
\begin{figure}
    \begin{center}
    \includegraphics[width=\columnwidth, clip]{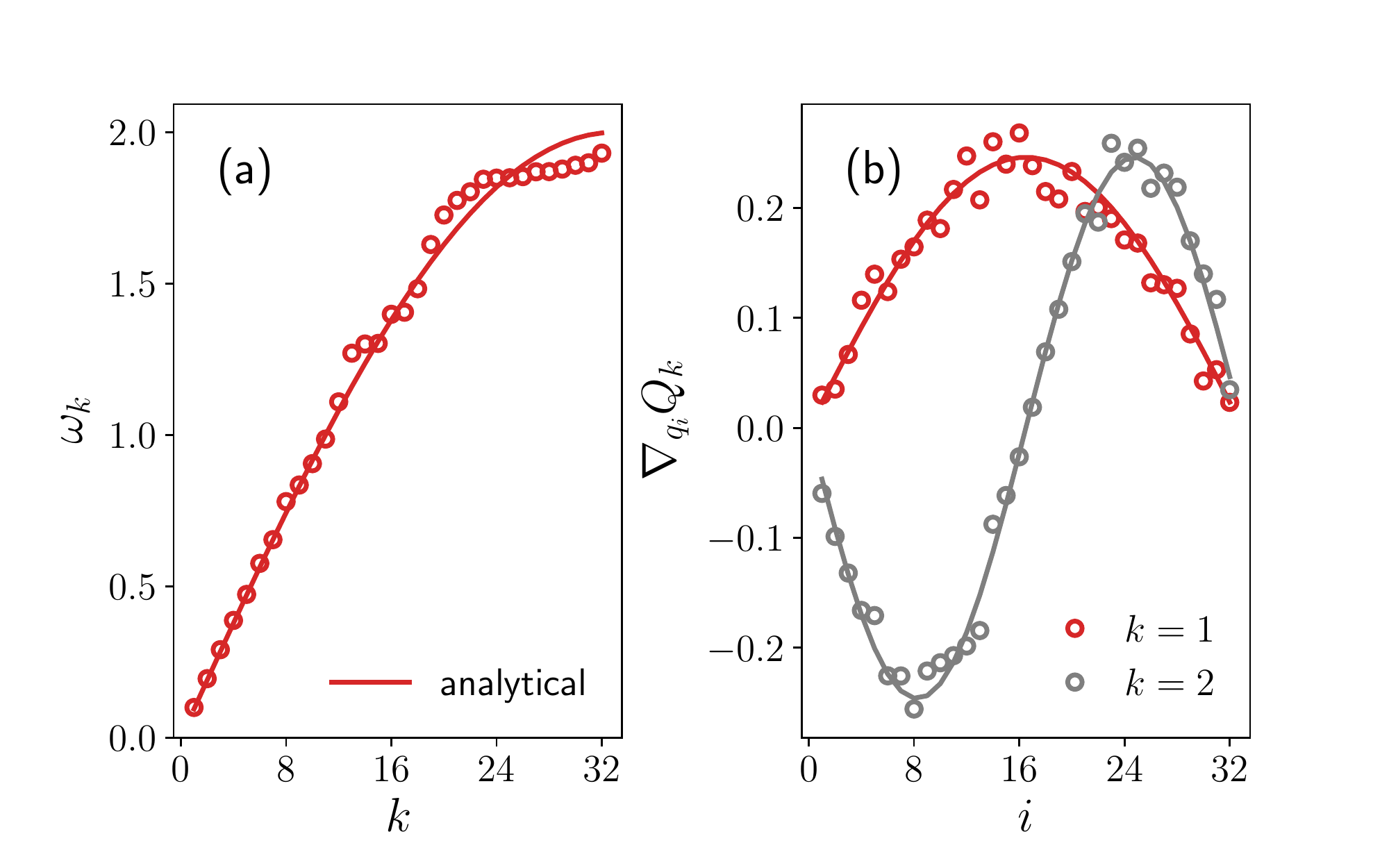} 
    \end{center}
    \caption{(a) The learned frequencies of harmonic chain present in Sec.~\ref{sec:harmonic} in the prior distribution [\Eq{eq:priorhamiltonian}] and the analytical normal mode frequency. (b) The Jacobian $\nabla_{q_i} Q_k$ of the two slowest collective coordinates with respect to the original coordinates. Solid lines are the analytical solution. }\label{fig:harmonicchain}
\end{figure}

Next, we consider a harmonic chain with the Hamiltonian $H = \frac{1}{2} \sum_{i=1}^n \left[ {p_i^2} +   (q_i-q_{i-1})^2 \right]$. We set $q_0 = q_{n+1}=0$ to fix both ends of the chain. 
The system can be readily diagonalized by finding the normal modes representation $H =\frac{1}{2}\sum_{k=1}^n(\dot{Q}_k^2 + \omega_k^2 Q_k^2)$, where $\omega_k = 2 \sin \left(\frac{\pi k}{2n+2}\right)$ is the normal modes frequency and $Q_k = \sqrt{\frac{2}{n+1}} \sum_{i=1}^n q_i \sin \left(\frac{ik \pi}{n+1}\right)$ is the normal coordinate. 

We train a neural point transformation with the variational loss \Eq{eq:pddloss} at the inverse temperature $\beta=1$. 
Figure~\ref{fig:harmonicchain}(a) shows learned frequencies in the latent space harmonic Hamiltonian \Eq{eq:priorhamiltonian} together with the analytical dispersion. 
The agreement is particularly good for the low frequencies which are populated by the canonical distribution at the given temperature. 
Moreover, we pick the two slowest coordinates  $Q_k$ and compute their Jacobians with respect to the physical variables $q_i$ as shown in Fig.~\ref{fig:harmonicchain}(b). 
The comparison shows that the neural canonical transformation nicely identifies slow collective modes of the system based on its Hamiltonian. On the other hand, the model is also able to learn these slow modes from data. In this simple case, modes with small frequency correspond to latent variables with large covariance, which could also be captured by the principal component analysis (PCA)~\cite{Pearson:1901gs}. 
%Using the neural point transformation \Eq{eq:coord} one can obtain similar results. 
%Setting up a linear symplectic transformation works enough for the harmonic chain. 
 
Having demonstrated that the neural point transformation reproduces conventional normal modes analysis and PCA for the harmonic chain, we move on to show the major strength of the present approach in extracting nonlinear slow modes. 
 
\subsection{Alanine Dipeptide} \label{sec:dipeptide}

\begin{figure}
    \begin{center}
    \includegraphics[width=\columnwidth, trim={0 0cm 0cm 0cm}, clip]{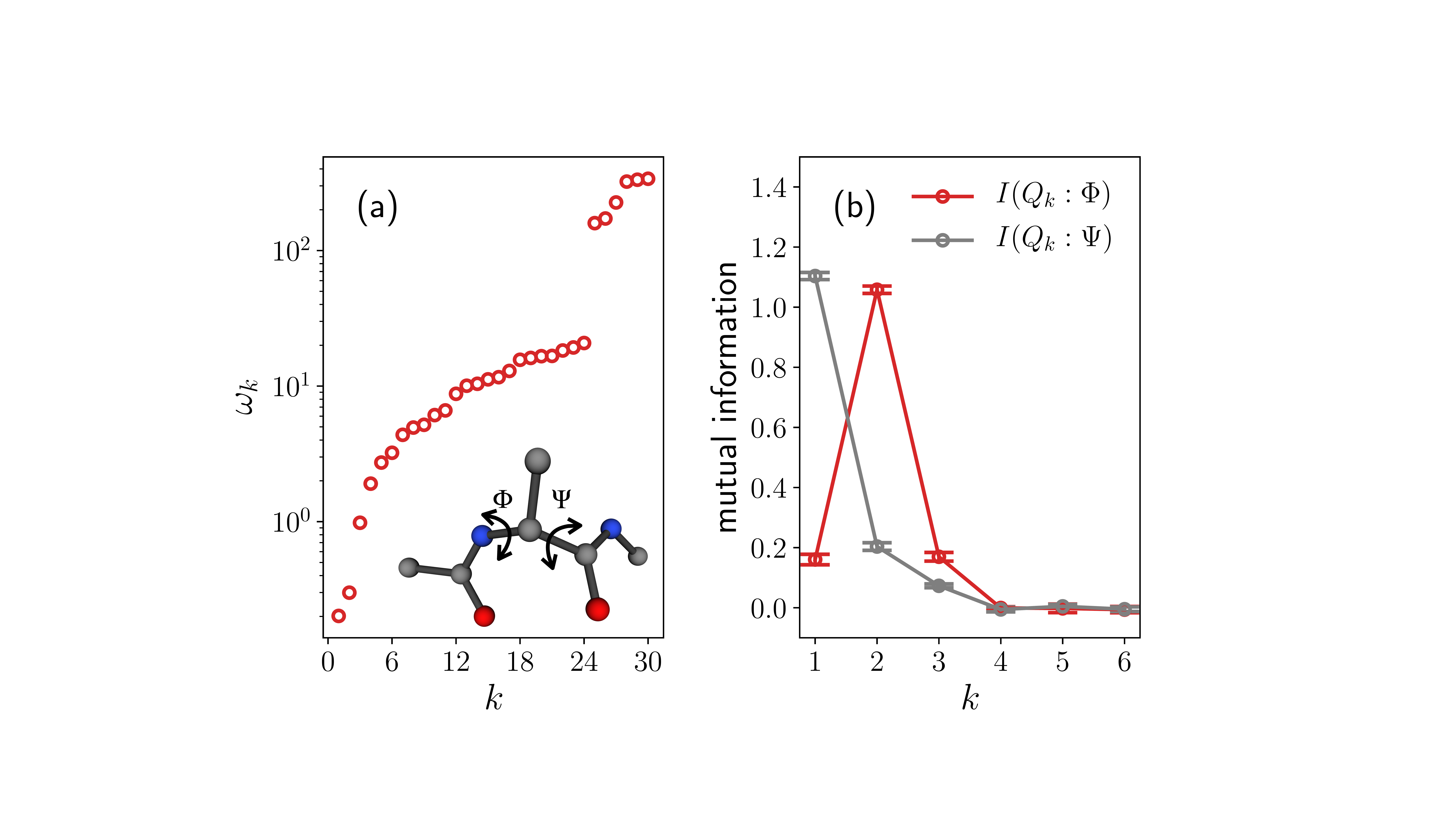} 
    \end{center}
    \caption{(a) The learned frequencies of alanine dipeptide presented in Sec.~\ref{sec:dipeptide} from the MD trajectories. Inset shows the molecule with slow torsion angles. (b) The mutual information between the few slowest modes and the torsion angles. 
    %(b,c,d) The MD trajectory samples projected on the plane of the three collective varaibles with the smallest frequencies versus the torsion angles.  
        }\label{fig:md}
\end{figure}

Proteins show rich dynamics with multiple emergent time scales. 
As one of the protein's building blocks, the alanine dipeptide is a standard benchmark problem.
Despite being a small organic molecule, the alanine dipeptide  shows nontrivial dynamics that deserve study. The backbone of alanine dipeptide contains $10$ heavy atoms with the SMILES representation \texttt{CC(=O)NC(C)C(=O)NC}. 
It is known that the two dihedral angles $\Phi$ and $\Psi$ which control the torsion of the molecule, as indicated in the inset of Fig.~\ref{fig:md}(a), are the key degrees of freedom which show slow dynamics.
% In the polypeptide it is these degrees of freedoms determines the hierarchical structure of protein.
% The usage of machine learning techniques has enjoy a great success in this field, except using flow-based model to achieve better sampling~\ref{}

Here, we train a neural canonical transformation to identify the slow modes of the alanine dipeptide molecule based on raw MD simulation data. 
For the training, we use the MD dataset \cite{Nuske2017, Wehmeyer2018} released at \cite{mdshare}. 
The dataset consists of $250~ns$ of Euclidean space trajectory of the $10$ heavy atoms in the alanine dipeptide at $300$K with an integration step of $2~fs$. 
Since the density estimation \Eq{eq:mleloss} requires the phase space data, we extent atom coordinates data to the phase space by sampling momenta from the Gaussian distribution whose variances depend on the atom masses and temperature. 
%The variances of the Gaussian distribution are determined by the atom masses and temperature. 
Note that for the phase space density estimation we randomly shuffle the trajectory data, thus we do not use any of the timeframe information in the training. 
We use the three MD independent trajectories at \cite{mdshare} for the training, validation, and testing, respectively. 
Each of them contains $250,000$ snapshots. 
We use the Adam optimizer~\cite{kingma2015adam} with a minibatch size $200$ and an initial learning rate $10^{-3}$ for training. 
We reduce the learning rate by a factor of $10$ if there is no improvement of the loss function on the validation set for $10$ training steps. 

%Therefore, the objective function becomes 
%\begin{equation}
%\mathcal{L}_\mathrm{MLE} = - \mathbb{E}_{\substack {p\sim \mathcal{N} \\ q \sim \mathcal{D}} } \left[\ln \rho(\boldsymbol{\eta})\right ].  
%\label{eq:agumented_mseloss} 
%\end{equation} 
Figure~\ref{fig:md}(a) shows the learned frequencies of the alanine dipeptide dataset, which spans a wide-scale and suggests the emergence of slow collective modes in the system. The frequency of the slowest modes is smaller than the fastest mode by more than three orders of magnitude. 
%Assuming the fastest mode is of the order of the integration time step, the time scale of the slowest modes correspond to more than a few thousands of femtoseconds. 
Moreover, there is a notable gap in the frequencies, which suggests a separation of the fast and slow modes in the system. 

\begin{figure*}
    \begin{center}
    \includegraphics[width=1.8\columnwidth, trim={0 0cm 0cm 0cm}, clip]{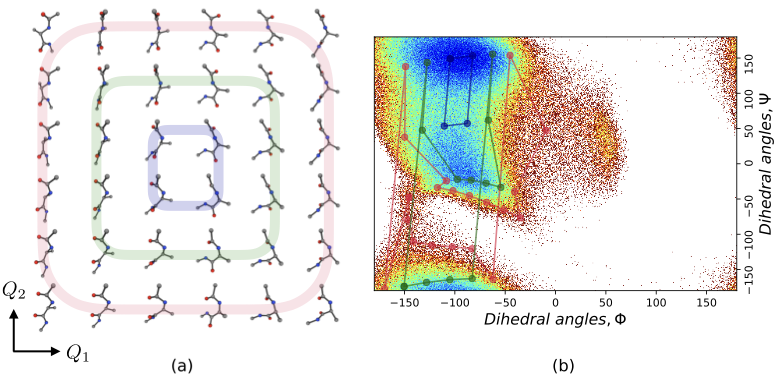} 
    \end{center}
    \caption{(a) Latent space interpolation in the plane of the two slowest collective coordinates ($Q_1,\ Q_2$) generates various molecule conformations. (b) Probability profile of the alanine dipeptide as a function of dihedral angles from generated samples of neural canonical transformation. 
     The paths corresponding to the path with the same color in (a). }\label{fig:dipeptide} 
\end{figure*}

We identify the latent coordinates with the smallest frequencies as the slowest nonlinear collective variables. To connect these learned collective variables to empirically known torsion angles, we estimate the mutual information \cite{Kraskov2004} between them and the two torsion angles $\Phi$ and $\Psi$ in Fig.~\ref{fig:md}(b).
%we plot the two slowest latent variables agains the known torsion angles $\Phi$ and $\Psi$ which corresponds to the dihedral angles of the $CNCC$ and $NCCN$ atoms respectively 
One sees that the first two latent coordinates show a significant correlation with $\Psi$ and $\Phi$, respectively. 
The mutual information between latent coordinates with larger frequencies and these two torsion angles decreases rapidly. 
Therefore, we conclude that the symplectic network has successfully identified the relevant slow modes which capture the low energy physics. Remarkably, this is done without having any access to the time information in the MD trajectory. 
This discovery highlights the usefulness of imposing the symplectic symmetry in the flow to turn statistical information into the dynamical one. We note that despite showing large mutual information, the learned two slowest coordinates does not exactly reproduce these torsion angles.
The reason being that the learning objective encourages to identify independent collective variables whose marginal distributions are independent Gaussians, while the marginal distributions of these torsion angles are clearly not. 
%In fact, this will not be the case since the learning objectivity encourages to identify independent collective variables whose marginal distribution are independent Gaussians.
Instead, this objective that favors independence allows us to gain better control over the identified latent variables, and thus offer advantages for practical purpose as we show next.
%For a practical purpose, identifying the low dimensional manifold spanned by nonlinear slow modes provide an equally good description of the collective behavior of the system. 

%Moreover, the latent vector does not simply become the Gaussian distribution since one did not match the target and the model distribution. 
%Although the two slowest modes identified by the network have significant statistical correlation with the torsion angles,

Since the normalizing flow model is a bijective generative model, one can directly map latent variables to molecular configurations. 
Figure~\ref{fig:dipeptide}(a) shows the generated molecular conformation by tuning the two slowest modes in the range $Q_k \in [-1/\omega_k, 1/\omega_k]$. One clearly sees that the two slowest variables control the global geometry of the molecule. Figure~\ref{fig:dipeptide}(b) shows generated samples in the two dimensional plane of torsion angles. The learned distribution is wider than the given dataset, which is a common feature of the density estimation using the normalizing flows~\cite{Papamakarios2019}. Figure~\ref{fig:dipeptide}(b) also shows that smooth paths in the spaces of learned slow latent variables may correspond to nontrivial paths in the torsion angles plane. The neural canonical transformation has learned a compact embedding molecular conformations in the learned latent space. 

\begin{figure}[h]
    \begin{center}
    \includegraphics[width=\columnwidth, trim={0 0cm 0cm 0cm}, clip]{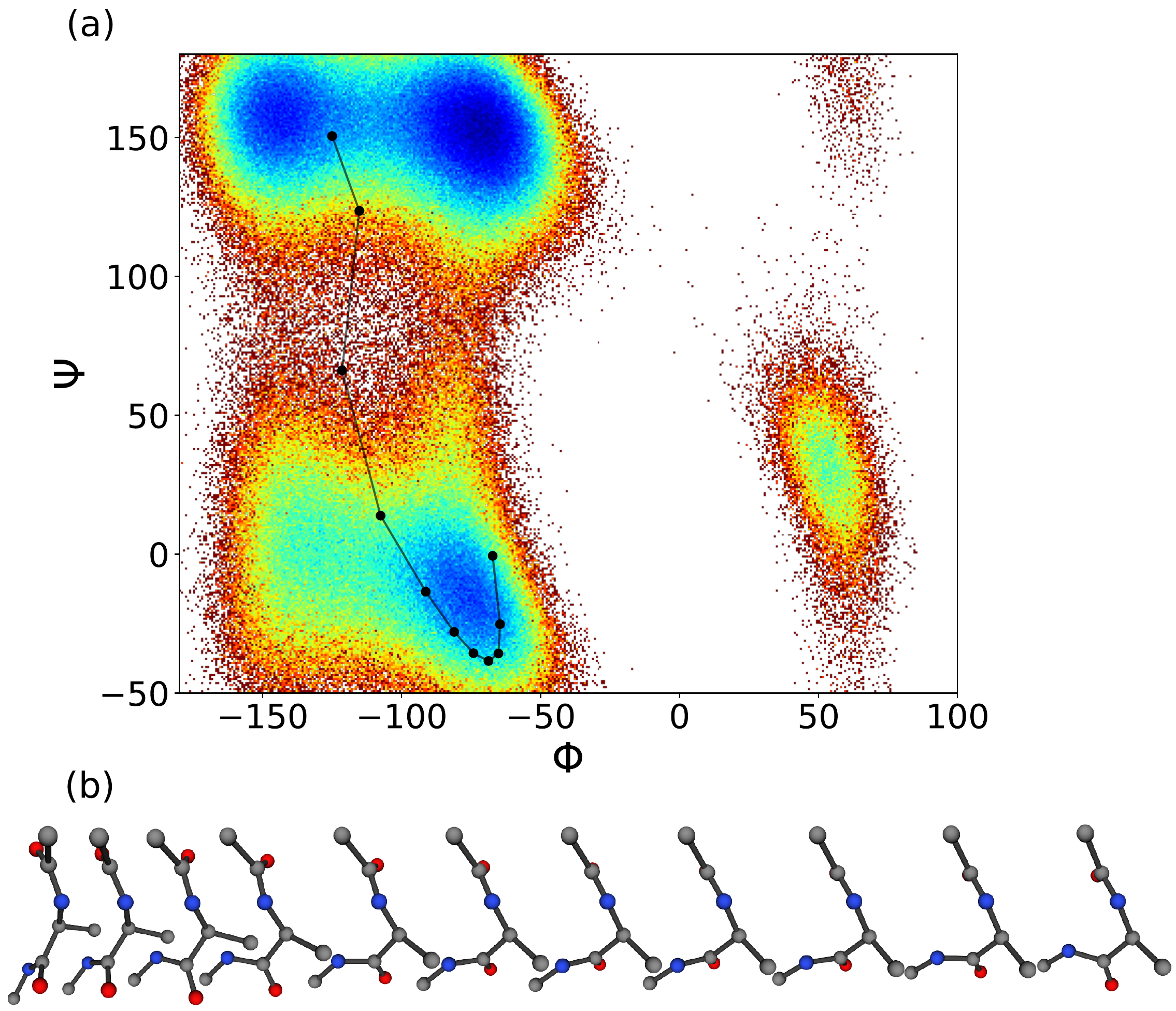} 
    \end{center}
    \caption{(a) A path from the molecular configuration at (\ang{-125},\ang{150}) to the configuration at (\ang{-75},\ang{0}) obtained by spherical linear interpolation of the two slowest latent variables. The background is the probability profile of the alanine dipeptide dataset on the plane of the dihedral angles. (b) The molecular conformation along the interpolation path.}\label{fig:sameEnInterp} 
\end{figure}

Having a compact latent representation of the molecular configurations allows one to design a smooth path between stable conformations by interpolating a few latent variables. Figure~\ref{fig:sameEnInterp} shows a path connecting two molecular conformations with the spherical linear interpolation (SLERP) of the two slowest variables~\cite{slerp}. The interpolation gives a path along the geodesic curve of the Gaussian distributed latent variables and thus avoids unlikely molecular conformations. As the figure shows, the interpolation yields a curved path in the torsion angle plane which avoids the low-density region that may occur with a naive interpolation in terms of atom coordinates or torsion angles. 

Moreover, thanks to the tractability of the normalizing flow model one have an exact likelihood on the path for the generated molecular unlike in the case of generating molecular conformation VAE or GAN. This quantitative access to the likelihood allows unbiased sampling of the configuration space with rejection sampling~\cite{Huang2017a, Liu2017f} or latent space Monte Carlo updates~\cite{Li2018z, BoltzmannGenerator}.

%This also allows the latent space interpolation to design a trajectory in original space, like in Figure~\ref{fig:sameEnInterp}(a). We can design a latent space trajectory that the latent space energy decreases smoothly at each step. This is easy to achieve in the latent for latent space is comprised of independent Gaussian distributions. One such algorithm is the spherical linear interpolation (SLERP)~\cite{slerp}. Moreover, we can even simplify the high dimensional interpolation into few-dimensional interpolation by only performing the interpolation in the $n_\mathrm{slow}$ slowest modes. Then, these trajectories can be mapped into physical space. As canonical transformation is a volume-preserve one. The energy change in the physical space should also be decreasing smoothly, as shown in Figure~\ref{fig:sameEnInterp}(a). Then, one can see from Figure~\ref{fig:sameEnInterp}(b) the simplified trajectories are close to the original one.

%Moreover, the symplectic structure in the flow model offers a clear indicator to pick up the slow variables. 

\subsection{MNIST handwritten digits}
Finally, we apply the neural canonical transformation to machine learning problems.
We consider the MNIST handwritten digits dataset, which contains $50000$ grayscale images of $28\times 28$ pixels. 
These images are divided into ten-digit classes. 
Treating the pixel values as coordinate variables, we can view the digits classes as stable conformations of a physical system~\footnote{See Ref.~\cite{Zhang2018r} for the preprocessing steps to map MNIST dataset to continuous variables.}. 
Similar to the dipeptide studied in the Sec.~\ref{sec:dipeptide}, one conjectures that the transition between conformations are slow, while the variations within the digits classes are the fast degrees of freedom. We assume each pixel has unity mass and augment the dataset with momenta. Then, we perform density estimation in the phase space to train a neural canonical transformation.  

\begin{figure}
    \begin{center}
    \includegraphics[width=\columnwidth, trim={0 0cm 0cm 0cm}, clip]{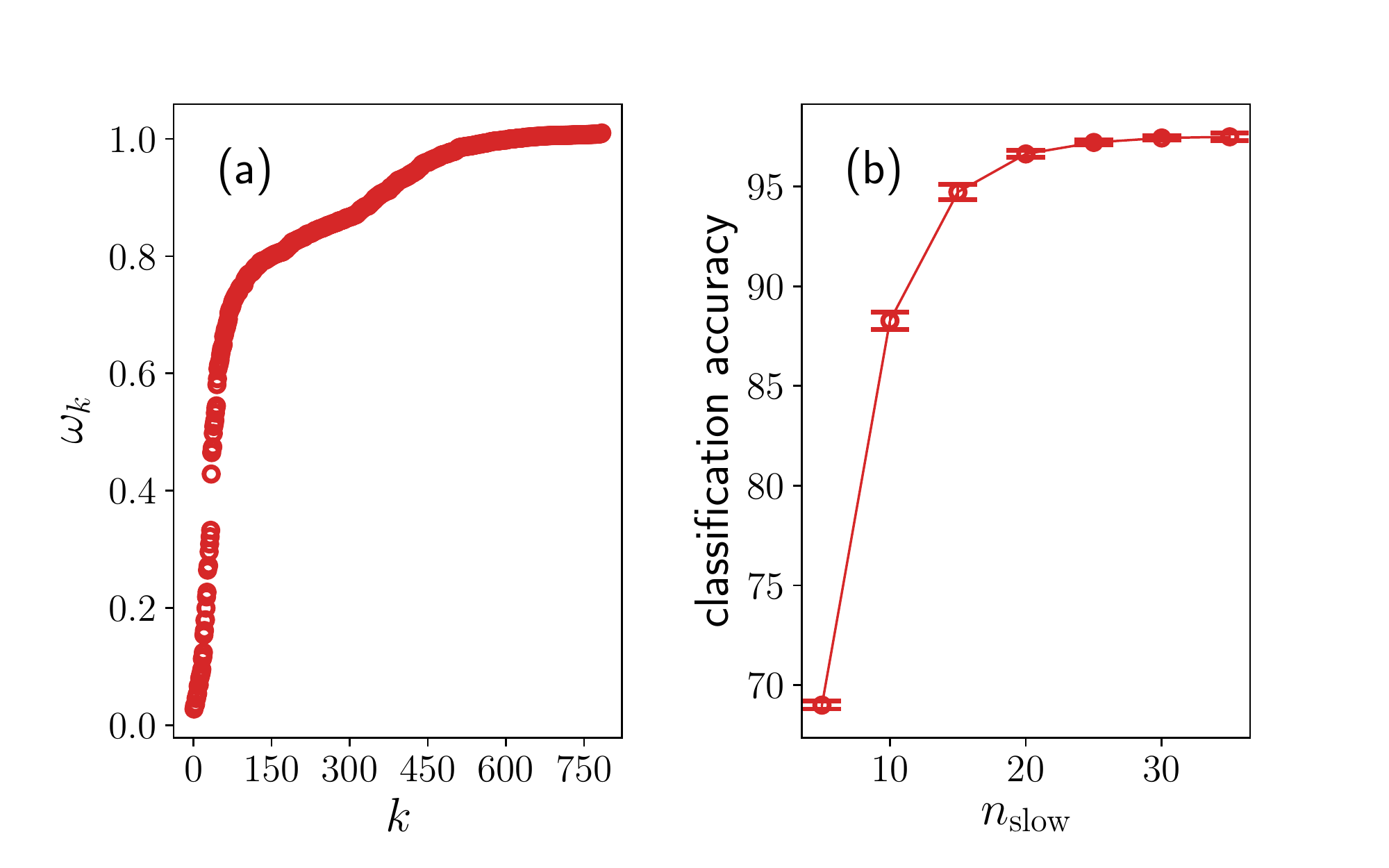} 
    \end{center}
    \caption{(a) The learned frequencies of the MNIST dataset. (b) Classification accuracy on the test dataset based on $n_\mathrm{slow}$ slowest modes.}\label{fig:MNIST_frequency}
\end{figure}

Figure~\ref{fig:MNIST_frequency}(a) shows the dispersion of the MNIST dataset, which contains a small portion of slow frequencies over all the variables.  
To show that these slow modes contain the relevant information of the digits classes, we pass only these slow modes to a multilayer perceptron classifier and perform supervised training.  
The classifier contains a single hidden of neurons with rectified linear units. 
The learned neural canonical transformation has its parameter fixed and works as a feature extractor. 
By varying the number of kept slow modes from $5$ up to $35$ out of the total $784$ dimensions, one sees that the classification accuracy on the test dataset quickly increases a plateau around $97\%$ as shown in Fig.~\ref{fig:MNIST_frequency}(b). Reaching high classification accuracy with only a few of the slow collective variables shows that they indeed capture digits class information. 

\begin{figure}[h!]
    \begin{center}
      \includegraphics[width=\columnwidth, trim={0 0cm 0cm 0cm}, clip]{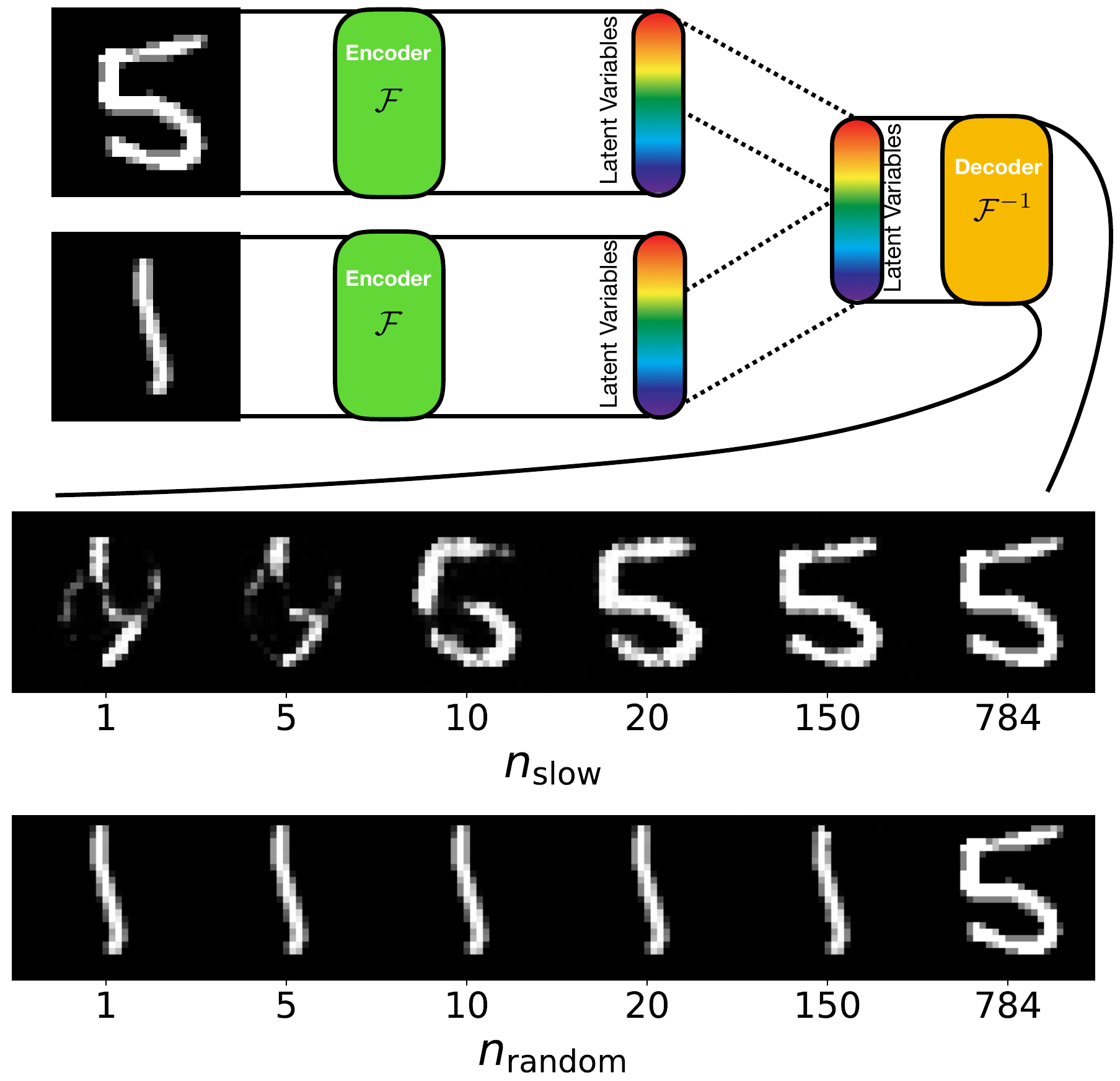} 
    \end{center}
    \caption{The coordinate transformation \Eq{eq:coord} maps the MNIST images to a latent representation with independent frequency modes. We concatenate slow latent variables of an image together with the fast modes of another image and then map the combined latent vector back to the image space with the inverse coordinates transformation. The bottom panel shows the same experiment by concatenating latent vectors at random.}
   \label{fig:interpolationMNIST}
  \end{figure}

We perform an additional experiment to directly show that the learned slow modes indeed capture the salient features of the MNIST images, i.e. the digit classes. As shown in the top panel of Fig.~\ref{fig:interpolationMNIST}, we take a pair of images from MNIST and map both to latent space. 
Then, we take $n_\mathrm{slow}$ slowest latent vector from one image and take the remaining ones from another image and concatenate them together to form a latent vector.
%the first $5$ up to all latent space slowest modes of the source image using the target image, the interpolation happens 
%i.e. interpolated latent variables are $(\underbrace{\frac{(R-r){\boldsymbol{s}}_{1}+r{\boldsymbol{t}}_{1}}{R}, \ldots, \frac{(R-r){\boldsymbol{s}}_{n_\mathrm{slow}}+r{\boldsymbol{t}}_{n_\mathrm{slow}}}{R}}_{n_\mathrm{slow}},\underbrace{\boldsymbol{s}_{n+1}, \ldots, \boldsymbol{s}_{784}}_{784-n_\mathrm{slow}})$, where $\boldsymbol{s}$ is the latent variables from start image and $\boldsymbol{t}$ is the target, $R$ is the overall interpolation steps and $r$ is the current step, $n$ is the first $n$ slowest modes.  
%\red{Why not also doing slerp here for consistence ?} \comment{For simplicity, we don't need SLERP to guarantee every configs in between is valid.} 
After mapping the concatenated latent vector back to the image space, we see that even using $20$ slowest modes one can already change the digit class. In comparison, if we perform the same experiment with randomly selected $n_\mathrm{random}$ modes without considering the frequency order, one sees that one needs to use a  much larger number of latent variables to make the transition of digit classes. 

In Appendix~\ref{app:compress} we perform conceptual compression~\cite{Gregor2016, Dinh2016a} using the slow modes learned by the neural canonical transformation.

\section{Discussions} \label{sec:discussions}

Neural canonical transformation extends a long-standing theoretical tool with symplectic normalizing flows. It provides a systematic way to simplify Hamiltonian dynamics and extract independent nonlinear slow modes of Hamiltonian systems and natural dataset. %The approach can extract dynamical information from statistical analysis in the phase space. Thus, neural canonical transformations 

Besides data analysis, the neural canonical transformation may open the door to address the sampling problem in MD simulations. As a bottom line, one is ready to employ the existing enhance sampling approaches~\cite{Laio2002, Valsson2014,Zhang2018RiD} with the learned slow modes as the collective variables. Since these collective variables are differentiable and exhibit slow dynamics, it fulfills the typical requirement of collective variables. Moreover, one can already sample feasible molecular configurations directly by exploiting the learned canonical transformation as a flow-based generative model. These samples would approximate the target distribution well if the generative model is trained well. One can further correct the sampling bias by using the generative model as proposals in the Markov chain Monte Carlo~\cite{Huang2017a, Liu2017f}. Lastly, one may also perform Monte Carlo sampling in the latent space and then map the latent vectors to the physical samples~\cite{Li2018z, BoltzmannGenerator}. These latent space Monte Carlo updates extend the enhanced sampling approach based on variable transformations~\cite{Zhu2002} to an adaptive setting.

It is also instructive to put the present approach in the contexts of probabilistic generative models~\cite{NormalizingFlow, Dinh2014, Dinh2016a, Kingma2018, Grathwohl2018}. Conventional generative models only concern about the statistical properties of data in the configuration space with coordinates only. While neural canonical transformations deal with phase space density. Therefore, they allow access to dynamical information by exploiting the symplectic structure in the transformation. Since the phase space density is factorized to the momenta and coordinates parts for separable Hamiltonians, the KL divergence can be written as the sum of two term for the KL divergences of the momenta and coordinates marginal distributions respectively. % in the momenta and the coordinate space respectively, where the transformation in the two spaces are related by the symplectic condition. 
The phase space KL divergence is lower bounded by the one in the configuration space~\cite{Salimans2014}. 
In this sense, one can view the momenta as auxiliary variables that regularizes the training of the neural coordinate transformations. 
%In addition, composing momenta together with the coordinates data in the density estimation can be viewed as a form of data augmentation. 

%disentangled representation
A pressing issue with the latent variable generative models is how to select a handful of most relevant latent variables after training. There have been various attempts to design hierarchical generative models~\cite{Dinh2016a, Chen2016b, higgins2017beta, Karras2018, Das2019} to capture global information of data with a few latent variables. 
The neural canonical transformation differs by selecting the collective variables according to the learned frequencies in the latent space. Therefore, one can assign a dynamical interpretation to the statistically learned latent representation. 
% symmetry representation
On the other hand, currently, we use a generic real NVP model to perform the coordinate transformation. 
Additional symmetry constraints like the invariance under translation, rotation, and permutation among identical particles, are useful for future applications~\cite{Rezende2019,Kohlerb}.
In this case, one may devise an  equivariant transformation by leveraging a symmetry-preserving energy model and using it to drive a gradient flow~\cite{zhang2018end,Zhang2018r}.
%However, it is argued that in general a disentangled latent representation would not be possible without inductive bias~\cite{Locatellob}.
%The canonical transformation formalism address this issue in the contexts of flow-based generative models. 

We remark that reaching a perfect harmonic Hamiltonian \Eq{eq:priorhamiltonian} in the latent space is generally not possible because it requires the original system to be integrable. 
In fact, a perfectly trained neural canonical transformation would reveal the invariant torus of integrable systems as shown in Ref.~\cite{Bondesan2019}. 
So we do not expect the periodic dynamics of the hidden variables under the prior harmonic Hamiltonian to replace the dynamics of the physical variables. But we do expect that the slow modes extracted from the procedure would be meaningful and useful for downstream tasks. More generally, the Kolmogorov-Arnold-Moser theory~\cite{dumas2014kam} shows that the phase space trajectory of nearly-integrable systems would only be deformed from quasiperiodic motions. Thus, we expect the neural canonical transformation would work well for systems with coherent collective motion. Along this line, the present approach may also be useful to study synchronization phenomena~\cite{RevModPhys.77.137}, where a collective motion emerges out of complex dynamical systems. Finally, extending the present work to more general time-dependent canonical transformations may give an even more powerful tool to study many-particle dynamical systems. 

%As the consequences, theoretically, the present approach only works in the integrable and near integrable system. 
%Compared to the traditional method to approach the characteristic spectrum, for example, the correlation function method in simulations of the liquid~\cite{allen2017computer}, the nonlinear property of the symplectic neural network leads to better performance and more generality.
%we could also apply the present approach to the map between the nonintegrable Hamiltonian system. 

%\red{respa sampling with separated time-scale}
%\red{dynamic signature of the phase transition ???}

\section{Acknowledgment}
We thank Austen Lamacraft, Masatoshi Imada, Yuan Wan, Pan Zhang, Yi-Zhuang You, Zi Cai, Lei-Han Tang, Yueshui Zhang, Dian Wu, Yantao Wu and Yixiao Chen for discussions. The work is supported by the Ministry of Science and Technology of China under Grants No. 2016YFA0300603 and No. 2016YFA0302400, the National Natural Science Foundation of China under Grant No. 11774398, the Strategic Priority Research Program of Chinese Academy of Sciences Grant No. XDB28000000, and the Computational Chemical Center: Chemistry in Solution and at Interfaces funded by the DoE under Award DE-SC0019394.
%The work is supported by the Ministry of Science and Technology of China under the Grant No.~2016YFA0300603 and  2016YFA0302400, the National Natural Science Foundation of China under Grant No.~11774398, and the Strategic Priority Research Program of Chinese Academy of Sciences Grant No.~XDB28000000.

%\bibliographystyle{apsrev4-1}
%\bibliography{/Users/wanglei/Documents/Papers2/papers}
\bibliography{refs,manual_refs}

%merlin.mbs apsrev4-1.bst 2010-07-25 4.21a (PWD, AO, DPC) hacked
%Control: key (0)
%Control: author (0) dotless jnrlst
%Control: editor formatted (1) identically to author
%Control: production of article title (0) allowed
%Control: page (1) range
%Control: year (0) verbatim
%Control: production of eprint (0) enabled
\begin{thebibliography}{87}%
\makeatletter
\providecommand \@ifxundefined [1]{%
 \@ifx{#1\undefined}
}%
\providecommand \@ifnum [1]{%
 \ifnum #1\expandafter \@firstoftwo
 \else \expandafter \@secondoftwo
 \fi
}%
\providecommand \@ifx [1]{%
 \ifx #1\expandafter \@firstoftwo
 \else \expandafter \@secondoftwo
 \fi
}%
\providecommand \natexlab [1]{#1}%
\providecommand \enquote  [1]{``#1''}%
\providecommand \bibnamefont  [1]{#1}%
\providecommand \bibfnamefont [1]{#1}%
\providecommand \citenamefont [1]{#1}%
\providecommand \href@noop [0]{\@secondoftwo}%
\providecommand \href [0]{\begingroup \@sanitize@url \@href}%
\providecommand \@href[1]{\@@startlink{#1}\@@href}%
\providecommand \@@href[1]{\endgroup#1\@@endlink}%
\providecommand \@sanitize@url [0]{\catcode `\\12\catcode `\$12\catcode
  `\&12\catcode `\#12\catcode `\^12\catcode `\_12\catcode `\%12\relax}%
\providecommand \@@startlink[1]{}%
\providecommand \@@endlink[0]{}%
\providecommand \url  [0]{\begingroup\@sanitize@url \@url }%
\providecommand \@url [1]{\endgroup\@href {#1}{\urlprefix }}%
\providecommand \urlprefix  [0]{URL }%
\providecommand \Eprint [0]{\href }%
\providecommand \doibase [0]{http://dx.doi.org/}%
\providecommand \selectlanguage [0]{\@gobble}%
\providecommand \bibinfo  [0]{\@secondoftwo}%
\providecommand \bibfield  [0]{\@secondoftwo}%
\providecommand \translation [1]{[#1]}%
\providecommand \BibitemOpen [0]{}%
\providecommand \bibitemStop [0]{}%
\providecommand \bibitemNoStop [0]{.\EOS\space}%
\providecommand \EOS [0]{\spacefactor3000\relax}%
\providecommand \BibitemShut  [1]{\csname bibitem#1\endcsname}%
\let\auto@bib@innerbib\@empty
%</preamble>
\bibitem [{\citenamefont {Arnold}(1989)}]{Arnold1989}%
  \BibitemOpen
  \bibfield  {author} {\bibinfo {author} {\bibfnamefont {V.~I.}\ \bibnamefont
  {Arnold}},\ }\href@noop {} {\emph {\bibinfo {title} {{Mathematical Methods of
  Classical Mechanics}}}}\ (\bibinfo  {publisher} {Springer},\ \bibinfo {year}
  {1989})\BibitemShut {NoStop}%
\bibitem [{\citenamefont {Liouville}(1838)}]{liouville1838note}%
  \BibitemOpen
  \bibfield  {author} {\bibinfo {author} {\bibfnamefont {Joseph}\ \bibnamefont
  {Liouville}},\ }\bibfield  {title} {\enquote {\bibinfo {title} {Note sur la
  th{\'e}orie de la variation des constantes arbitraires.}}\ }\href@noop {}
  {\bibfield  {journal} {\bibinfo  {journal} {Journal de math{\'e}matiques
  pures et appliqu{\'e}es}\ ,\ \bibinfo {pages} {342--349}} (\bibinfo {year}
  {1838})}\BibitemShut {NoStop}%
\bibitem [{\citenamefont {Feng}\ and\ \citenamefont {Qin}(2011)}]{Feng2011}%
  \BibitemOpen
  \bibfield  {author} {\bibinfo {author} {\bibfnamefont {Kang}\ \bibnamefont
  {Feng}}\ and\ \bibinfo {author} {\bibfnamefont {Mengzhao}\ \bibnamefont
  {Qin}},\ }\href {\doibase 10.1007/978-3-642-01777-3} {\emph {\bibinfo {title}
  {{Symplectic Geometric Algorithms for Hamiltonian Systems}}}}\ (\bibinfo
  {publisher} {Springer},\ \bibinfo {year} {2011})\BibitemShut {NoStop}%
\bibitem [{\citenamefont {Shaw}\ \emph {et~al.}(2010)\citenamefont {Shaw},
  \citenamefont {Maragakis}, \citenamefont {Lindorff-Larsen}, \citenamefont
  {Piana}, \citenamefont {Dror}, \citenamefont {Eastwood}, \citenamefont
  {Bank}, \citenamefont {Jumper}, \citenamefont {Salmon}, \citenamefont
  {Shan},\ and\ \citenamefont {Wriggers}}]{Lindorff-Larsen2010}%
  \BibitemOpen
  \bibfield  {author} {\bibinfo {author} {\bibfnamefont {David~E}\ \bibnamefont
  {Shaw}}, \bibinfo {author} {\bibfnamefont {Paul}\ \bibnamefont {Maragakis}},
  \bibinfo {author} {\bibfnamefont {Kresten}\ \bibnamefont {Lindorff-Larsen}},
  \bibinfo {author} {\bibfnamefont {Stefano}\ \bibnamefont {Piana}}, \bibinfo
  {author} {\bibfnamefont {Ron~O}\ \bibnamefont {Dror}}, \bibinfo {author}
  {\bibfnamefont {Michael~P}\ \bibnamefont {Eastwood}}, \bibinfo {author}
  {\bibfnamefont {Joseph~A}\ \bibnamefont {Bank}}, \bibinfo {author}
  {\bibfnamefont {John~M}\ \bibnamefont {Jumper}}, \bibinfo {author}
  {\bibfnamefont {John~K}\ \bibnamefont {Salmon}}, \bibinfo {author}
  {\bibfnamefont {Yibing}\ \bibnamefont {Shan}}, \ and\ \bibinfo {author}
  {\bibfnamefont {Willy}\ \bibnamefont {Wriggers}},\ }\bibfield  {title}
  {\enquote {\bibinfo {title} {{Atomic-level characterization of the structural
  dynamics of proteins}},}\ }\href {\doibase 10.1126/science.1187409}
  {\bibfield  {journal} {\bibinfo  {journal} {Science}\ }\textbf {\bibinfo
  {volume} {330}},\ \bibinfo {pages} {341--346} (\bibinfo {year}
  {2010})}\BibitemShut {NoStop}%
\bibitem [{\citenamefont {No{\'{e}}}(2018)}]{Noe}%
  \BibitemOpen
  \bibfield  {author} {\bibinfo {author} {\bibfnamefont {Frank}\ \bibnamefont
  {No{\'{e}}}},\ }\bibfield  {title} {\enquote {\bibinfo {title} {{Machine
  Learning for Molecular Dynamics on Long Timescales}},}\ }\href
  {http://arxiv.org/abs/1812.07669} {\  (\bibinfo {year} {2018})},\ \Eprint
  {http://arxiv.org/abs/1812.07669} {arXiv:1812.07669} \BibitemShut {NoStop}%
\bibitem [{\citenamefont {Wang}\ \emph {et~al.}(2019)\citenamefont {Wang},
  \citenamefont {Ribeiro},\ and\ \citenamefont {Tiwary}}]{Wang2019}%
  \BibitemOpen
  \bibfield  {author} {\bibinfo {author} {\bibfnamefont {Yihang}\ \bibnamefont
  {Wang}}, \bibinfo {author} {\bibfnamefont {Joao Marcelo~Lamim}\ \bibnamefont
  {Ribeiro}}, \ and\ \bibinfo {author} {\bibfnamefont {Pratyush}\ \bibnamefont
  {Tiwary}},\ }\bibfield  {title} {\enquote {\bibinfo {title} {{Machine
  learning approaches for analyzing and enhancing molecular dynamics
  simulations}},}\ }\href {http://arxiv.org/abs/1909.11748} {\  (\bibinfo
  {year} {2019})},\ \Eprint {http://arxiv.org/abs/1909.11748}
  {arXiv:1909.11748} \BibitemShut {NoStop}%
\bibitem [{\citenamefont {Molgedey}\ and\ \citenamefont
  {Schuster}(1994)}]{Molgedey1994}%
  \BibitemOpen
  \bibfield  {author} {\bibinfo {author} {\bibfnamefont {L.}~\bibnamefont
  {Molgedey}}\ and\ \bibinfo {author} {\bibfnamefont {H.~G.}\ \bibnamefont
  {Schuster}},\ }\bibfield  {title} {\enquote {\bibinfo {title} {{Separation of
  a Mixture of Independent Signals Using Time Delayed Correlations}},}\ }\href
  {https://doi.org/10.1103/PhysRevLett.72.3634} {\bibfield  {journal} {\bibinfo
   {journal} {Phys. Rev. Lett.}\ }\textbf {\bibinfo {volume} {72}},\ \bibinfo
  {pages} {3634} (\bibinfo {year} {1994})}\BibitemShut {NoStop}%
\bibitem [{\citenamefont {{Belouchrani}}\ \emph {et~al.}(1997)\citenamefont
  {{Belouchrani}}, \citenamefont {{Abed-Meraim}}, \citenamefont {{Cardoso}},\
  and\ \citenamefont {{Moulines}}}]{Klaus-Robert97}%
  \BibitemOpen
  \bibfield  {author} {\bibinfo {author} {\bibfnamefont {A.}~\bibnamefont
  {{Belouchrani}}}, \bibinfo {author} {\bibfnamefont {K.}~\bibnamefont
  {{Abed-Meraim}}}, \bibinfo {author} {\bibfnamefont {J.~.}\ \bibnamefont
  {{Cardoso}}}, \ and\ \bibinfo {author} {\bibfnamefont {E.}~\bibnamefont
  {{Moulines}}},\ }\bibfield  {title} {\enquote {\bibinfo {title} {A blind
  source separation technique using second-order statistics},}\ }\href
  {\doibase 10.1109/78.554307} {\bibfield  {journal} {\bibinfo  {journal} {IEEE
  Transactions on Signal Processing}\ }\textbf {\bibinfo {volume} {45}},\
  \bibinfo {pages} {434--444} (\bibinfo {year} {1997})}\BibitemShut {NoStop}%
\bibitem [{\citenamefont {Ziehe}\ and\ \citenamefont
  {M{\"u}ller}(1998)}]{Klaus-Robert98}%
  \BibitemOpen
  \bibfield  {author} {\bibinfo {author} {\bibfnamefont {Andreas}\ \bibnamefont
  {Ziehe}}\ and\ \bibinfo {author} {\bibfnamefont {Klaus-Robert}\ \bibnamefont
  {M{\"u}ller}},\ }\href@noop {} {\emph {\bibinfo {title} {ICANN 98}}},\ edited
  by\ \bibinfo {editor} {\bibfnamefont {Lars}\ \bibnamefont {Niklasson}},
  \bibinfo {editor} {\bibfnamefont {Mikael}\ \bibnamefont {Bod{\'e}n}}, \ and\
  \bibinfo {editor} {\bibfnamefont {Tom}\ \bibnamefont {Ziemke}}\ (\bibinfo
  {publisher} {Springer London},\ \bibinfo {address} {London},\ \bibinfo {year}
  {1998})\ pp.\ \bibinfo {pages} {675--680}\BibitemShut {NoStop}%
\bibitem [{\citenamefont {P{\'{e}}rez-Hern{\'{a}}ndez}\ \emph
  {et~al.}(2013)\citenamefont {P{\'{e}}rez-Hern{\'{a}}ndez}, \citenamefont
  {Paul}, \citenamefont {Giorgino}, \citenamefont {{De Fabritiis}},\ and\
  \citenamefont {No{\'{e}}}}]{Perez-Hernandez2013}%
  \BibitemOpen
  \bibfield  {author} {\bibinfo {author} {\bibfnamefont {Guillermo}\
  \bibnamefont {P{\'{e}}rez-Hern{\'{a}}ndez}}, \bibinfo {author} {\bibfnamefont
  {Fabian}\ \bibnamefont {Paul}}, \bibinfo {author} {\bibfnamefont {Toni}\
  \bibnamefont {Giorgino}}, \bibinfo {author} {\bibfnamefont {Gianni}\
  \bibnamefont {{De Fabritiis}}}, \ and\ \bibinfo {author} {\bibfnamefont
  {Frank}\ \bibnamefont {No{\'{e}}}},\ }\bibfield  {title} {\enquote {\bibinfo
  {title} {{Identification of slow molecular order parameters for Markov model
  construction}},}\ }\href {https://doi.org/10.1063/1.4811489} {\bibfield
  {journal} {\bibinfo  {journal} {J. Chem. Phys.}\ }\textbf {\bibinfo {volume}
  {139}} (\bibinfo {year} {2013})}\BibitemShut {NoStop}%
\bibitem [{\citenamefont {Schwantes}\ and\ \citenamefont
  {Pande}(2013)}]{Schwantes2013}%
  \BibitemOpen
  \bibfield  {author} {\bibinfo {author} {\bibfnamefont {Christian~R.}\
  \bibnamefont {Schwantes}}\ and\ \bibinfo {author} {\bibfnamefont {Vijay~S.}\
  \bibnamefont {Pande}},\ }\bibfield  {title} {\enquote {\bibinfo {title}
  {{Improvements in Markov State Model construction reveal many non-native
  interactions in the folding of NTL9}},}\ }\href {\doibase 10.1021/ct300878a}
  {\bibfield  {journal} {\bibinfo  {journal} {J. Chem. Theory Comput.}\
  }\textbf {\bibinfo {volume} {9}},\ \bibinfo {pages} {2000--2009} (\bibinfo
  {year} {2013})}\BibitemShut {NoStop}%
\bibitem [{\citenamefont {Schmid}(2010)}]{Schmid2010}%
  \BibitemOpen
  \bibfield  {author} {\bibinfo {author} {\bibfnamefont {Peter~J.}\
  \bibnamefont {Schmid}},\ }\bibfield  {title} {\enquote {\bibinfo {title}
  {{Dynamic mode decomposition of numerical and experimental data}},}\ }\href
  {\doibase 10.1017/S0022112010001217} {\bibfield  {journal} {\bibinfo
  {journal} {J. Fluid Mech.}\ }\textbf {\bibinfo {volume} {656}},\ \bibinfo
  {pages} {5--28} (\bibinfo {year} {2010})}\BibitemShut {NoStop}%
\bibitem [{\citenamefont {Klus}\ \emph {et~al.}(2018)\citenamefont {Klus},
  \citenamefont {N{\"{u}}ske}, \citenamefont {Koltai}, \citenamefont {Wu},
  \citenamefont {Kevrekidis}, \citenamefont {Sch{\"{u}}tte},\ and\
  \citenamefont {No{\'{e}}}}]{Klus2018}%
  \BibitemOpen
  \bibfield  {author} {\bibinfo {author} {\bibfnamefont {Stefan}\ \bibnamefont
  {Klus}}, \bibinfo {author} {\bibfnamefont {Feliks}\ \bibnamefont
  {N{\"{u}}ske}}, \bibinfo {author} {\bibfnamefont {P{\'{e}}ter}\ \bibnamefont
  {Koltai}}, \bibinfo {author} {\bibfnamefont {Hao}\ \bibnamefont {Wu}},
  \bibinfo {author} {\bibfnamefont {Ioannis}\ \bibnamefont {Kevrekidis}},
  \bibinfo {author} {\bibfnamefont {Christof}\ \bibnamefont {Sch{\"{u}}tte}}, \
  and\ \bibinfo {author} {\bibfnamefont {Frank}\ \bibnamefont {No{\'{e}}}},\
  }\bibfield  {title} {\enquote {\bibinfo {title} {{Data-Driven Model Reduction
  and Transfer Operator Approximation}},}\ }\href {\doibase
  10.1007/s00332-017-9437-7} {\bibfield  {journal} {\bibinfo  {journal} {J.
  Nonlinear Sci.}\ }\textbf {\bibinfo {volume} {28}},\ \bibinfo {pages}
  {985--1010} (\bibinfo {year} {2018})}\BibitemShut {NoStop}%
\bibitem [{\citenamefont {Sch{\"o}lkopf}\ \emph {et~al.}(1998)\citenamefont
  {Sch{\"o}lkopf}, \citenamefont {Smola},\ and\ \citenamefont
  {M{\"u}ller}}]{Klaus-Robert98Nonl}%
  \BibitemOpen
  \bibfield  {author} {\bibinfo {author} {\bibfnamefont {Bernhard}\
  \bibnamefont {Sch{\"o}lkopf}}, \bibinfo {author} {\bibfnamefont {Alexander}\
  \bibnamefont {Smola}}, \ and\ \bibinfo {author} {\bibfnamefont
  {Klaus-Robert}\ \bibnamefont {M{\"u}ller}},\ }\bibfield  {title} {\enquote
  {\bibinfo {title} {Nonlinear component analysis as a kernel eigenvalue
  problem},}\ }\href {\doibase 10.1162/089976698300017467} {\bibfield
  {journal} {\bibinfo  {journal} {Neural Computation}\ }\textbf {\bibinfo
  {volume} {10}},\ \bibinfo {pages} {1299--1319} (\bibinfo {year} {1998})},\
  \Eprint {http://arxiv.org/abs/https://doi.org/10.1162/089976698300017467}
  {https://doi.org/10.1162/089976698300017467} \BibitemShut {NoStop}%
\bibitem [{\citenamefont {Harmeling}\ \emph {et~al.}(2003)\citenamefont
  {Harmeling}, \citenamefont {Ziehe}, \citenamefont {Kawanabe},\ and\
  \citenamefont {M{\"u}ller}}]{Klaus-Robert03}%
  \BibitemOpen
  \bibfield  {author} {\bibinfo {author} {\bibfnamefont {Stefan}\ \bibnamefont
  {Harmeling}}, \bibinfo {author} {\bibfnamefont {Andreas}\ \bibnamefont
  {Ziehe}}, \bibinfo {author} {\bibfnamefont {Motoaki}\ \bibnamefont
  {Kawanabe}}, \ and\ \bibinfo {author} {\bibfnamefont {Klaus-Robert}\
  \bibnamefont {M{\"u}ller}},\ }\bibfield  {title} {\enquote {\bibinfo {title}
  {Kernel-based nonlinear blind source separation},}\ }\href {\doibase
  10.1162/089976603765202677} {\bibfield  {journal} {\bibinfo  {journal}
  {Neural Computation}\ }\textbf {\bibinfo {volume} {15}},\ \bibinfo {pages}
  {1089--1124} (\bibinfo {year} {2003})},\ \Eprint
  {http://arxiv.org/abs/https://doi.org/10.1162/089976603765202677}
  {https://doi.org/10.1162/089976603765202677} \BibitemShut {NoStop}%
\bibitem [{\citenamefont {Mardt}\ \emph {et~al.}(2018)\citenamefont {Mardt},
  \citenamefont {Pasquali}, \citenamefont {Wu},\ and\ \citenamefont
  {No{\'{e}}}}]{Mardt2018}%
  \BibitemOpen
  \bibfield  {author} {\bibinfo {author} {\bibfnamefont {Andreas}\ \bibnamefont
  {Mardt}}, \bibinfo {author} {\bibfnamefont {Luca}\ \bibnamefont {Pasquali}},
  \bibinfo {author} {\bibfnamefont {Hao}\ \bibnamefont {Wu}}, \ and\ \bibinfo
  {author} {\bibfnamefont {Frank}\ \bibnamefont {No{\'{e}}}},\ }\bibfield
  {title} {\enquote {\bibinfo {title} {{VAMPnets for deep learning of molecular
  kinetics}},}\ }\href {\doibase 10.1038/s41467-017-02388-1} {\bibfield
  {journal} {\bibinfo  {journal} {Nat. Commun.}\ }\textbf {\bibinfo {volume}
  {9}},\ \bibinfo {pages} {5} (\bibinfo {year} {2018})},\ \Eprint
  {http://arxiv.org/abs/1710.06012} {arXiv:1710.06012} \BibitemShut {NoStop}%
\bibitem [{\citenamefont {Wehmeyer}\ and\ \citenamefont
  {No{\'{e}}}(2018)}]{Wehmeyer2018}%
  \BibitemOpen
  \bibfield  {author} {\bibinfo {author} {\bibfnamefont {Christoph}\
  \bibnamefont {Wehmeyer}}\ and\ \bibinfo {author} {\bibfnamefont {Frank}\
  \bibnamefont {No{\'{e}}}},\ }\bibfield  {title} {\enquote {\bibinfo {title}
  {{Time-lagged autoencoders: Deep learning of slow collective variables for
  molecular kinetics}},}\ }\href {https://doi.org/10.1063/1.5011399} {\bibfield
   {journal} {\bibinfo  {journal} {J. Chem. Phys.}\ }\textbf {\bibinfo {volume}
  {148}},\ \bibinfo {pages} {241703} (\bibinfo {year} {2018})}\BibitemShut
  {NoStop}%
\bibitem [{\citenamefont {Hern{\'{a}}ndez}\ \emph {et~al.}(2018)\citenamefont
  {Hern{\'{a}}ndez}, \citenamefont {Wayment-Steele}, \citenamefont {Sultan},
  \citenamefont {Husic},\ and\ \citenamefont {Pande}}]{Hernandez2018a}%
  \BibitemOpen
  \bibfield  {author} {\bibinfo {author} {\bibfnamefont {Carlos~X.}\
  \bibnamefont {Hern{\'{a}}ndez}}, \bibinfo {author} {\bibfnamefont
  {Hannah~K.}\ \bibnamefont {Wayment-Steele}}, \bibinfo {author} {\bibfnamefont
  {Mohammad~M.}\ \bibnamefont {Sultan}}, \bibinfo {author} {\bibfnamefont
  {Brooke~E.}\ \bibnamefont {Husic}}, \ and\ \bibinfo {author} {\bibfnamefont
  {Vijay~S.}\ \bibnamefont {Pande}},\ }\bibfield  {title} {\enquote {\bibinfo
  {title} {{Variational encoding of complex dynamics}},}\ }\href {\doibase
  10.1103/PhysRevE.97.062412} {\bibfield  {journal} {\bibinfo  {journal} {Phys.
  Rev. E}\ }\textbf {\bibinfo {volume} {97}},\ \bibinfo {pages} {062412}
  (\bibinfo {year} {2018})}\BibitemShut {NoStop}%
\bibitem [{\citenamefont {Sultan}\ and\ \citenamefont {Pande}(2018)}]{Sultan}%
  \BibitemOpen
  \bibfield  {author} {\bibinfo {author} {\bibfnamefont {Mohammad~M.}\
  \bibnamefont {Sultan}}\ and\ \bibinfo {author} {\bibfnamefont {Vijay~S.}\
  \bibnamefont {Pande}},\ }\bibfield  {title} {\enquote {\bibinfo {title}
  {{Decision functions from supervised machine learning algorithms as
  collective variables for accelerating molecular simulations}},}\ }\href@noop
  {} {\  (\bibinfo {year} {2018})},\ \Eprint {http://arxiv.org/abs/1802.10510}
  {arXiv:1802.10510} \BibitemShut {NoStop}%
\bibitem [{\citenamefont {Lusch}\ \emph {et~al.}(2018)\citenamefont {Lusch},
  \citenamefont {Kutz},\ and\ \citenamefont {Brunton}}]{Lusch2018}%
  \BibitemOpen
  \bibfield  {author} {\bibinfo {author} {\bibfnamefont {Bethany}\ \bibnamefont
  {Lusch}}, \bibinfo {author} {\bibfnamefont {J.~Nathan}\ \bibnamefont {Kutz}},
  \ and\ \bibinfo {author} {\bibfnamefont {Steven~L.}\ \bibnamefont
  {Brunton}},\ }\bibfield  {title} {\enquote {\bibinfo {title} {{Deep learning
  for universal linear embeddings of nonlinear dynamics}},}\ }\href {\doibase
  10.1038/s41467-018-07210-0} {\bibfield  {journal} {\bibinfo  {journal} {Nat.
  Commun.}\ }\textbf {\bibinfo {volume} {9}} (\bibinfo {year} {2018}),\
  10.1038/s41467-018-07210-0}\BibitemShut {NoStop}%
\bibitem [{\citenamefont {Wiskott}\ and\ \citenamefont
  {Sejnowski}(2002)}]{Wiskott2002a}%
  \BibitemOpen
  \bibfield  {author} {\bibinfo {author} {\bibfnamefont {Laurenz}\ \bibnamefont
  {Wiskott}}\ and\ \bibinfo {author} {\bibfnamefont {Terrence~J.}\ \bibnamefont
  {Sejnowski}},\ }\bibfield  {title} {\enquote {\bibinfo {title} {{Slow feature
  analysis: Unsupervised learning of invariances}},}\ }\href {\doibase
  10.1162/089976602317318938} {\bibfield  {journal} {\bibinfo  {journal}
  {Neural Comput.}\ }\textbf {\bibinfo {volume} {14}},\ \bibinfo {pages}
  {715--770} (\bibinfo {year} {2002})}\BibitemShut {NoStop}%
\bibitem [{\citenamefont {Pfau}\ \emph {et~al.}(2018)\citenamefont {Pfau},
  \citenamefont {Petersen}, \citenamefont {Agarwal}, \citenamefont {Barrett},\
  and\ \citenamefont {Stachenfeld}}]{Pfau2018a}%
  \BibitemOpen
  \bibfield  {author} {\bibinfo {author} {\bibfnamefont {David}\ \bibnamefont
  {Pfau}}, \bibinfo {author} {\bibfnamefont {Stig}\ \bibnamefont {Petersen}},
  \bibinfo {author} {\bibfnamefont {Ashish}\ \bibnamefont {Agarwal}}, \bibinfo
  {author} {\bibfnamefont {David G.~T.}\ \bibnamefont {Barrett}}, \ and\
  \bibinfo {author} {\bibfnamefont {Kimberly~L.}\ \bibnamefont {Stachenfeld}},\
  }\bibfield  {title} {\enquote {\bibinfo {title} {{Spectral Inference
  Networks: Unifying Deep and Spectral Learning}},}\ }\href
  {http://arxiv.org/abs/1806.02215} {\  (\bibinfo {year} {2018})},\ \Eprint
  {http://arxiv.org/abs/1806.02215} {arXiv:1806.02215} \BibitemShut {NoStop}%
\bibitem [{\citenamefont {Kobyzev}\ \emph {et~al.}(2019)\citenamefont
  {Kobyzev}, \citenamefont {Prince},\ and\ \citenamefont
  {Brubaker}}]{Kobyzev2019}%
  \BibitemOpen
  \bibfield  {author} {\bibinfo {author} {\bibfnamefont {Ivan}\ \bibnamefont
  {Kobyzev}}, \bibinfo {author} {\bibfnamefont {Simon}\ \bibnamefont {Prince}},
  \ and\ \bibinfo {author} {\bibfnamefont {Marcus~A.}\ \bibnamefont
  {Brubaker}},\ }\bibfield  {title} {\enquote {\bibinfo {title} {{Normalizing
  Flows: Introduction and Ideas}},}\ }\href {http://arxiv.org/abs/1908.09257}
  {\  (\bibinfo {year} {2019})},\ \Eprint {http://arxiv.org/abs/1908.09257}
  {arXiv:1908.09257} \BibitemShut {NoStop}%
\bibitem [{\citenamefont {Papamakarios}\ \emph {et~al.}(2019)\citenamefont
  {Papamakarios}, \citenamefont {Nalisnick}, \citenamefont {Rezende},
  \citenamefont {Mohamed},\ and\ \citenamefont
  {Lakshminarayanan}}]{Papamakarios2019}%
  \BibitemOpen
  \bibfield  {author} {\bibinfo {author} {\bibfnamefont {George}\ \bibnamefont
  {Papamakarios}}, \bibinfo {author} {\bibfnamefont {Eric}\ \bibnamefont
  {Nalisnick}}, \bibinfo {author} {\bibfnamefont {Danilo~Jimenez}\ \bibnamefont
  {Rezende}}, \bibinfo {author} {\bibfnamefont {Shakir}\ \bibnamefont
  {Mohamed}}, \ and\ \bibinfo {author} {\bibfnamefont {Balaji}\ \bibnamefont
  {Lakshminarayanan}},\ }\bibfield  {title} {\enquote {\bibinfo {title}
  {{Normalizing Flows for Probabilistic Modeling and Inference}},}\ }\href
  {http://arxiv.org/abs/1912.02762} {\  (\bibinfo {year} {2019})},\ \Eprint
  {http://arxiv.org/abs/1912.02762} {arXiv:1912.02762} \BibitemShut {NoStop}%
\bibitem [{\citenamefont {Mattheakis}\ \emph {et~al.}(2019)\citenamefont
  {Mattheakis}, \citenamefont {Protopapas}, \citenamefont {Sondak},
  \citenamefont {{Di Giovanni}},\ and\ \citenamefont {Kaxiras}}]{Mattheakis}%
  \BibitemOpen
  \bibfield  {author} {\bibinfo {author} {\bibfnamefont {M}~\bibnamefont
  {Mattheakis}}, \bibinfo {author} {\bibfnamefont {P}~\bibnamefont
  {Protopapas}}, \bibinfo {author} {\bibfnamefont {D}~\bibnamefont {Sondak}},
  \bibinfo {author} {\bibfnamefont {M}~\bibnamefont {{Di Giovanni}}}, \ and\
  \bibinfo {author} {\bibfnamefont {E}~\bibnamefont {Kaxiras}},\ }\bibfield
  {title} {\enquote {\bibinfo {title} {{Physical Symmetries Embedded in Neural
  Networks}},}\ }\href {https://arxiv.org/pdf/1904.08991.pdf} {\bibfield
  {journal} {\bibinfo  {journal} {arXiv}\ } (\bibinfo {year} {2019})},\ \Eprint
  {http://arxiv.org/abs/1904.08991v1} {arXiv:1904.08991v1} \BibitemShut
  {NoStop}%
\bibitem [{\citenamefont {Greydanus}\ \emph {et~al.}(2019)\citenamefont
  {Greydanus}, \citenamefont {Dzamba},\ and\ \citenamefont
  {Yosinski}}]{Greydanus2019}%
  \BibitemOpen
  \bibfield  {author} {\bibinfo {author} {\bibfnamefont {Sam}\ \bibnamefont
  {Greydanus}}, \bibinfo {author} {\bibfnamefont {Misko}\ \bibnamefont
  {Dzamba}}, \ and\ \bibinfo {author} {\bibfnamefont {Jason}\ \bibnamefont
  {Yosinski}},\ }\bibfield  {title} {\enquote {\bibinfo {title} {{Hamiltonian
  Neural Networks}},}\ }\href {http://arxiv.org/abs/1906.01563} {\  (\bibinfo
  {year} {2019})},\ \Eprint {http://arxiv.org/abs/1906.01563}
  {arXiv:1906.01563} \BibitemShut {NoStop}%
\bibitem [{\citenamefont {Sanchez-Gonzalez}\ \emph {et~al.}(2019)\citenamefont
  {Sanchez-Gonzalez}, \citenamefont {Bapst}, \citenamefont {Cranmer},\ and\
  \citenamefont {Battaglia}}]{Sanchez-Gonzalez2019}%
  \BibitemOpen
  \bibfield  {author} {\bibinfo {author} {\bibfnamefont {Alvaro}\ \bibnamefont
  {Sanchez-Gonzalez}}, \bibinfo {author} {\bibfnamefont {Victor}\ \bibnamefont
  {Bapst}}, \bibinfo {author} {\bibfnamefont {Kyle}\ \bibnamefont {Cranmer}}, \
  and\ \bibinfo {author} {\bibfnamefont {Peter}\ \bibnamefont {Battaglia}},\
  }\bibfield  {title} {\enquote {\bibinfo {title} {{Hamiltonian Graph Networks
  with ODE Integrators}},}\ }\href {http://arxiv.org/abs/1909.12790} {\
  (\bibinfo {year} {2019})},\ \Eprint {http://arxiv.org/abs/1909.12790}
  {arXiv:1909.12790} \BibitemShut {NoStop}%
\bibitem [{\citenamefont {Zhong}\ \emph {et~al.}(2019)\citenamefont {Zhong},
  \citenamefont {Dey},\ and\ \citenamefont {Chakraborty}}]{Zhong2019}%
  \BibitemOpen
  \bibfield  {author} {\bibinfo {author} {\bibfnamefont {Yaofeng~Desmond}\
  \bibnamefont {Zhong}}, \bibinfo {author} {\bibfnamefont {Biswadip}\
  \bibnamefont {Dey}}, \ and\ \bibinfo {author} {\bibfnamefont {Amit}\
  \bibnamefont {Chakraborty}},\ }\bibfield  {title} {\enquote {\bibinfo {title}
  {{Symplectic ODE-Net: Learning Hamiltonian Dynamics with Control}},}\ }\href
  {http://arxiv.org/abs/1909.12077} {\  (\bibinfo {year} {2019})},\ \Eprint
  {http://arxiv.org/abs/1909.12077} {arXiv:1909.12077} \BibitemShut {NoStop}%
\bibitem [{\citenamefont {Chen}\ \emph {et~al.}(2019)\citenamefont {Chen},
  \citenamefont {Zhang}, \citenamefont {Arjovsky},\ and\ \citenamefont
  {Bottou}}]{Chen2019}%
  \BibitemOpen
  \bibfield  {author} {\bibinfo {author} {\bibfnamefont {Zhengdao}\
  \bibnamefont {Chen}}, \bibinfo {author} {\bibfnamefont {Jianyu}\ \bibnamefont
  {Zhang}}, \bibinfo {author} {\bibfnamefont {Martin}\ \bibnamefont
  {Arjovsky}}, \ and\ \bibinfo {author} {\bibfnamefont {L{\'{e}}on}\
  \bibnamefont {Bottou}},\ }\bibfield  {title} {\enquote {\bibinfo {title}
  {{Symplectic Recurrent Neural Networks}},}\ }\href
  {http://arxiv.org/abs/1909.13334} {\  (\bibinfo {year} {2019})},\ \Eprint
  {http://arxiv.org/abs/1909.13334} {arXiv:1909.13334} \BibitemShut {NoStop}%
\bibitem [{\citenamefont {Rezende}\ \emph {et~al.}(2019)\citenamefont
  {Rezende}, \citenamefont {Racani{\`{e}}re}, \citenamefont {Higgins},\ and\
  \citenamefont {Toth}}]{Rezende2019}%
  \BibitemOpen
  \bibfield  {author} {\bibinfo {author} {\bibfnamefont {Danilo~J.}\
  \bibnamefont {Rezende}}, \bibinfo {author} {\bibfnamefont {S{\'{e}}bastien}\
  \bibnamefont {Racani{\`{e}}re}}, \bibinfo {author} {\bibfnamefont {Irina}\
  \bibnamefont {Higgins}}, \ and\ \bibinfo {author} {\bibfnamefont {Peter}\
  \bibnamefont {Toth}},\ }\bibfield  {title} {\enquote {\bibinfo {title}
  {{Equivariant Hamiltonian Flows}},}\ }\href {http://arxiv.org/abs/1909.13739}
  {\  (\bibinfo {year} {2019})},\ \Eprint {http://arxiv.org/abs/1909.13739}
  {arXiv:1909.13739} \BibitemShut {NoStop}%
\bibitem [{\citenamefont {Toth}\ \emph {et~al.}(2019)\citenamefont {Toth},
  \citenamefont {Rezende}, \citenamefont {Jaegle}, \citenamefont
  {Racani{\`{e}}re}, \citenamefont {Botev},\ and\ \citenamefont
  {Higgins}}]{Toth2019}%
  \BibitemOpen
  \bibfield  {author} {\bibinfo {author} {\bibfnamefont {Peter}\ \bibnamefont
  {Toth}}, \bibinfo {author} {\bibfnamefont {Danilo~Jimenez}\ \bibnamefont
  {Rezende}}, \bibinfo {author} {\bibfnamefont {Andrew}\ \bibnamefont
  {Jaegle}}, \bibinfo {author} {\bibfnamefont {S{\'{e}}bastien}\ \bibnamefont
  {Racani{\`{e}}re}}, \bibinfo {author} {\bibfnamefont {Aleksandar}\
  \bibnamefont {Botev}}, \ and\ \bibinfo {author} {\bibfnamefont {Irina}\
  \bibnamefont {Higgins}},\ }\bibfield  {title} {\enquote {\bibinfo {title}
  {{Hamiltonian Generative Networks}},}\ }\href
  {http://arxiv.org/abs/1909.13789} {\  (\bibinfo {year} {2019})},\ \Eprint
  {http://arxiv.org/abs/1909.13789} {arXiv:1909.13789} \BibitemShut {NoStop}%
\bibitem [{\citenamefont {Bondesan}\ and\ \citenamefont
  {Lamacraft}(2019)}]{Bondesan2019}%
  \BibitemOpen
  \bibfield  {author} {\bibinfo {author} {\bibfnamefont {Roberto}\ \bibnamefont
  {Bondesan}}\ and\ \bibinfo {author} {\bibfnamefont {Austen}\ \bibnamefont
  {Lamacraft}},\ }\bibfield  {title} {\enquote {\bibinfo {title} {{Learning
  Symmetries of Classical Integrable Systems}},}\ }\href
  {http://arxiv.org/abs/1906.04645} {\  (\bibinfo {year} {2019})},\ \Eprint
  {http://arxiv.org/abs/1906.04645} {arXiv:1906.04645} \BibitemShut {NoStop}%
\bibitem [{\citenamefont {Behler}\ and\ \citenamefont
  {Parrinello}(2007)}]{behler2007generalized}%
  \BibitemOpen
  \bibfield  {author} {\bibinfo {author} {\bibfnamefont {J{\"o}rg}\
  \bibnamefont {Behler}}\ and\ \bibinfo {author} {\bibfnamefont {Michele}\
  \bibnamefont {Parrinello}},\ }\bibfield  {title} {\enquote {\bibinfo {title}
  {Generalized neural-network representation of high-dimensional
  potential-energy surfaces},}\ }\href@noop {} {\bibfield  {journal} {\bibinfo
  {journal} {Physical Review Letters}\ }\textbf {\bibinfo {volume} {98}},\
  \bibinfo {pages} {146401} (\bibinfo {year} {2007})}\BibitemShut {NoStop}%
\bibitem [{\citenamefont {Sch{\"u}tt}\ \emph {et~al.}(2017)\citenamefont
  {Sch{\"u}tt}, \citenamefont {Arbabzadah}, \citenamefont {Chmiela},
  \citenamefont {M{\"u}ller},\ and\ \citenamefont
  {Tkatchenko}}]{schutt2017quantum}%
  \BibitemOpen
  \bibfield  {author} {\bibinfo {author} {\bibfnamefont {Kristof~T}\
  \bibnamefont {Sch{\"u}tt}}, \bibinfo {author} {\bibfnamefont {Farhad}\
  \bibnamefont {Arbabzadah}}, \bibinfo {author} {\bibfnamefont {Stefan}\
  \bibnamefont {Chmiela}}, \bibinfo {author} {\bibfnamefont {Klaus~R}\
  \bibnamefont {M{\"u}ller}}, \ and\ \bibinfo {author} {\bibfnamefont
  {Alexandre}\ \bibnamefont {Tkatchenko}},\ }\bibfield  {title} {\enquote
  {\bibinfo {title} {Quantum-chemical insights from deep tensor neural
  networks},}\ }\href@noop {} {\bibfield  {journal} {\bibinfo  {journal}
  {Nature Communications}\ }\textbf {\bibinfo {volume} {8}},\ \bibinfo {pages}
  {13890} (\bibinfo {year} {2017})}\BibitemShut {NoStop}%
\bibitem [{\citenamefont {Han}\ \emph {et~al.}(2018)\citenamefont {Han},
  \citenamefont {Zhang}, \citenamefont {Car},\ and\ \citenamefont
  {E}}]{han2017deep}%
  \BibitemOpen
  \bibfield  {author} {\bibinfo {author} {\bibfnamefont {Jiequn}\ \bibnamefont
  {Han}}, \bibinfo {author} {\bibfnamefont {Linfeng}\ \bibnamefont {Zhang}},
  \bibinfo {author} {\bibfnamefont {Roberto}\ \bibnamefont {Car}}, \ and\
  \bibinfo {author} {\bibfnamefont {Weinan}\ \bibnamefont {E}},\ }\bibfield
  {title} {\enquote {\bibinfo {title} {Deep potential: a general representation
  of a many-body potential energy surface},}\ }\href@noop {} {\bibfield
  {journal} {\bibinfo  {journal} {Communications in Computational Physics}\
  }\textbf {\bibinfo {volume} {23}},\ \bibinfo {pages} {629--639} (\bibinfo
  {year} {2018})}\BibitemShut {NoStop}%
\bibitem [{\citenamefont {Zhang}\ \emph
  {et~al.}(2018{\natexlab{a}})\citenamefont {Zhang}, \citenamefont {Han},
  \citenamefont {Wang}, \citenamefont {Car},\ and\ \citenamefont
  {E}}]{zhang2018deep}%
  \BibitemOpen
  \bibfield  {author} {\bibinfo {author} {\bibfnamefont {Linfeng}\ \bibnamefont
  {Zhang}}, \bibinfo {author} {\bibfnamefont {Jiequn}\ \bibnamefont {Han}},
  \bibinfo {author} {\bibfnamefont {Han}\ \bibnamefont {Wang}}, \bibinfo
  {author} {\bibfnamefont {Roberto}\ \bibnamefont {Car}}, \ and\ \bibinfo
  {author} {\bibfnamefont {Weinan}\ \bibnamefont {E}},\ }\bibfield  {title}
  {\enquote {\bibinfo {title} {Deep potential molecular dynamics: A scalable
  model with the accuracy of quantum mechanics},}\ }\href@noop {} {\bibfield
  {journal} {\bibinfo  {journal} {Physical Review Letters}\ }\textbf {\bibinfo
  {volume} {120}},\ \bibinfo {pages} {143001} (\bibinfo {year}
  {2018}{\natexlab{a}})}\BibitemShut {NoStop}%
\bibitem [{\citenamefont {Zhang}\ \emph
  {et~al.}(2018{\natexlab{b}})\citenamefont {Zhang}, \citenamefont {Han},
  \citenamefont {Wang}, \citenamefont {Saidi}, \citenamefont {Car},\ and\
  \citenamefont {E}}]{zhang2018end}%
  \BibitemOpen
  \bibfield  {author} {\bibinfo {author} {\bibfnamefont {Linfeng}\ \bibnamefont
  {Zhang}}, \bibinfo {author} {\bibfnamefont {Jiequn}\ \bibnamefont {Han}},
  \bibinfo {author} {\bibfnamefont {Han}\ \bibnamefont {Wang}}, \bibinfo
  {author} {\bibfnamefont {Wissam~A}\ \bibnamefont {Saidi}}, \bibinfo {author}
  {\bibfnamefont {Roberto}\ \bibnamefont {Car}}, \ and\ \bibinfo {author}
  {\bibfnamefont {Weinan}\ \bibnamefont {E}},\ }\bibfield  {title} {\enquote
  {\bibinfo {title} {End-to-end symmetry preserving inter-atomic potential
  energy model for finite and extended systems},}\ }in\ \href@noop {} {\emph
  {\bibinfo {booktitle} {Advances of the Neural Information Processing Systems
  (NIPS)}}}\ (\bibinfo {year} {2018})\BibitemShut {NoStop}%
\bibitem [{\citenamefont {Gregor}\ \emph
  {et~al.}(2016{\natexlab{a}})\citenamefont {Gregor}, \citenamefont {Besse},
  \citenamefont {Rezende}, \citenamefont {Danihelka},\ and\ \citenamefont
  {Wierstra}}]{Chmiela2018}%
  \BibitemOpen
  \bibfield  {author} {\bibinfo {author} {\bibfnamefont {Karol}\ \bibnamefont
  {Gregor}}, \bibinfo {author} {\bibfnamefont {Frederic}\ \bibnamefont
  {Besse}}, \bibinfo {author} {\bibfnamefont {Danilo~Jimenez}\ \bibnamefont
  {Rezende}}, \bibinfo {author} {\bibfnamefont {Ivo}\ \bibnamefont
  {Danihelka}}, \ and\ \bibinfo {author} {\bibfnamefont {Daan}\ \bibnamefont
  {Wierstra}},\ }\bibfield  {title} {\enquote {\bibinfo {title} {{Towards
  Conceptual Compression}},}\ }\href {http://arxiv.org/abs/1604.08772} {\
  (\bibinfo {year} {2016}{\natexlab{a}})},\ \Eprint
  {http://arxiv.org/abs/1604.08772} {arXiv:1604.08772} \BibitemShut {NoStop}%
\bibitem [{\citenamefont {Chmiela}\ \emph {et~al.}(2017)\citenamefont
  {Chmiela}, \citenamefont {Tkatchenko}, \citenamefont {Sauceda}, \citenamefont
  {Poltavsky}, \citenamefont {Sch{\"u}tt},\ and\ \citenamefont
  {M{\"u}ller}}]{Chmielae1603015}%
  \BibitemOpen
  \bibfield  {author} {\bibinfo {author} {\bibfnamefont {Stefan}\ \bibnamefont
  {Chmiela}}, \bibinfo {author} {\bibfnamefont {Alexandre}\ \bibnamefont
  {Tkatchenko}}, \bibinfo {author} {\bibfnamefont {Huziel~E.}\ \bibnamefont
  {Sauceda}}, \bibinfo {author} {\bibfnamefont {Igor}\ \bibnamefont
  {Poltavsky}}, \bibinfo {author} {\bibfnamefont {Kristof~T.}\ \bibnamefont
  {Sch{\"u}tt}}, \ and\ \bibinfo {author} {\bibfnamefont {Klaus-Robert}\
  \bibnamefont {M{\"u}ller}},\ }\bibfield  {title} {\enquote {\bibinfo {title}
  {Machine learning of accurate energy-conserving molecular force fields},}\
  }\href {\doibase 10.1126/sciadv.1603015} {\bibfield  {journal} {\bibinfo
  {journal} {Science Advances}\ }\textbf {\bibinfo {volume} {3}} (\bibinfo
  {year} {2017}),\ 10.1126/sciadv.1603015},\ \Eprint
  {http://arxiv.org/abs/https://advances.sciencemag.org/content/3/5/e1603015.full.pdf}
  {https://advances.sciencemag.org/content/3/5/e1603015.full.pdf} \BibitemShut
  {NoStop}%
\bibitem [{\citenamefont {Goodfellow}(2016)}]{Goodfellow2016a}%
  \BibitemOpen
  \bibfield  {author} {\bibinfo {author} {\bibfnamefont {Ian}\ \bibnamefont
  {Goodfellow}},\ }\bibfield  {title} {\enquote {\bibinfo {title} {{NIPS 2016
  Tutorial: Generative Adversarial Networks}},}\ }\href
  {http://arxiv.org/abs/1701.00160} {\  (\bibinfo {year} {2016})},\ \Eprint
  {http://arxiv.org/abs/1701.00160} {arXiv:1701.00160} \BibitemShut {NoStop}%
\bibitem [{\citenamefont {Goodfellow}\ \emph {et~al.}(2014)\citenamefont
  {Goodfellow}, \citenamefont {Pouget-Abadie}, \citenamefont {Mirza},
  \citenamefont {Xu}, \citenamefont {Warde-Farley}, \citenamefont {Ozair},
  \citenamefont {Courville},\ and\ \citenamefont {Bengio}}]{Goodfellow2014a}%
  \BibitemOpen
  \bibfield  {author} {\bibinfo {author} {\bibfnamefont {Ian~J.}\ \bibnamefont
  {Goodfellow}}, \bibinfo {author} {\bibfnamefont {Jean}\ \bibnamefont
  {Pouget-Abadie}}, \bibinfo {author} {\bibfnamefont {Mehdi}\ \bibnamefont
  {Mirza}}, \bibinfo {author} {\bibfnamefont {Bing}\ \bibnamefont {Xu}},
  \bibinfo {author} {\bibfnamefont {David}\ \bibnamefont {Warde-Farley}},
  \bibinfo {author} {\bibfnamefont {Sherjil}\ \bibnamefont {Ozair}}, \bibinfo
  {author} {\bibfnamefont {Aaron}\ \bibnamefont {Courville}}, \ and\ \bibinfo
  {author} {\bibfnamefont {Yoshua}\ \bibnamefont {Bengio}},\ }\bibfield
  {title} {\enquote {\bibinfo {title} {{Generative Adversarial Networks}},}\
  }\href {http://arxiv.org/abs/1406.2661} {\  (\bibinfo {year} {2014})},\
  \Eprint {http://arxiv.org/abs/1406.2661} {arXiv:1406.2661} \BibitemShut
  {NoStop}%
\bibitem [{\citenamefont {Kingma}\ and\ \citenamefont
  {Welling}(2013)}]{Kingma2013}%
  \BibitemOpen
  \bibfield  {author} {\bibinfo {author} {\bibfnamefont {Diederik~P}\
  \bibnamefont {Kingma}}\ and\ \bibinfo {author} {\bibfnamefont {Max}\
  \bibnamefont {Welling}},\ }\bibfield  {title} {\enquote {\bibinfo {title}
  {{Auto-Encoding Variational Bayes}},}\ }\href
  {http://arxiv.org/abs/1312.6114} {\  (\bibinfo {year} {2013})},\ \Eprint
  {http://arxiv.org/abs/1312.6114} {arXiv:1312.6114} \BibitemShut {NoStop}%
\bibitem [{\citenamefont {Rezende}\ and\ \citenamefont
  {Mohamed}(2015)}]{NormalizingFlow}%
  \BibitemOpen
  \bibfield  {author} {\bibinfo {author} {\bibfnamefont {Danilo~Jimenez}\
  \bibnamefont {Rezende}}\ and\ \bibinfo {author} {\bibfnamefont {Shakir}\
  \bibnamefont {Mohamed}},\ }\bibfield  {title} {\enquote {\bibinfo {title}
  {{Variational Inference with Normalizing Flows}},}\ }\href
  {http://arxiv.org/abs/1505.05770} {\  (\bibinfo {year} {2015})},\ \Eprint
  {http://arxiv.org/abs/1505.05770} {arXiv:1505.05770} \BibitemShut {NoStop}%
\bibitem [{\citenamefont {Dinh}\ \emph {et~al.}(2014)\citenamefont {Dinh},
  \citenamefont {Krueger},\ and\ \citenamefont {Bengio}}]{Dinh2014}%
  \BibitemOpen
  \bibfield  {author} {\bibinfo {author} {\bibfnamefont {Laurent}\ \bibnamefont
  {Dinh}}, \bibinfo {author} {\bibfnamefont {David}\ \bibnamefont {Krueger}}, \
  and\ \bibinfo {author} {\bibfnamefont {Yoshua}\ \bibnamefont {Bengio}},\
  }\bibfield  {title} {\enquote {\bibinfo {title} {{NICE: Non-linear
  Independent Components Estimation}},}\ }\href
  {http://arxiv.org/abs/1410.8516} {\  (\bibinfo {year} {2014})},\ \Eprint
  {http://arxiv.org/abs/1410.8516} {arXiv:1410.8516} \BibitemShut {NoStop}%
\bibitem [{\citenamefont {Dinh}\ \emph {et~al.}(2016)\citenamefont {Dinh},
  \citenamefont {Sohl-Dickstein},\ and\ \citenamefont {Bengio}}]{Dinh2016a}%
  \BibitemOpen
  \bibfield  {author} {\bibinfo {author} {\bibfnamefont {Laurent}\ \bibnamefont
  {Dinh}}, \bibinfo {author} {\bibfnamefont {Jascha}\ \bibnamefont
  {Sohl-Dickstein}}, \ and\ \bibinfo {author} {\bibfnamefont {Samy}\
  \bibnamefont {Bengio}},\ }\bibfield  {title} {\enquote {\bibinfo {title}
  {{Density estimation using Real NVP}},}\ }\href
  {http://arxiv.org/abs/1605.08803} {\  (\bibinfo {year} {2016})},\ \Eprint
  {http://arxiv.org/abs/1605.08803} {arXiv:1605.08803} \BibitemShut {NoStop}%
\bibitem [{\citenamefont {Kingma}\ and\ \citenamefont
  {Dhariwal}(2018)}]{Kingma2018}%
  \BibitemOpen
  \bibfield  {author} {\bibinfo {author} {\bibfnamefont {Diederik~P}\
  \bibnamefont {Kingma}}\ and\ \bibinfo {author} {\bibfnamefont {Prafulla}\
  \bibnamefont {Dhariwal}},\ }\bibfield  {title} {\enquote {\bibinfo {title}
  {{Glow: Generative Flow with Invertible 1x1 Convolutions}},}\ }\href
  {http://arxiv.org/abs/1807.03039} {\  (\bibinfo {year} {2018})},\ \Eprint
  {http://arxiv.org/abs/1807.03039} {arXiv:1807.03039} \BibitemShut {NoStop}%
\bibitem [{\citenamefont {Grathwohl}\ \emph {et~al.}(2018)\citenamefont
  {Grathwohl}, \citenamefont {Chen}, \citenamefont {Bettencourt}, \citenamefont
  {Sutskever},\ and\ \citenamefont {Duvenaud}}]{Grathwohl2018}%
  \BibitemOpen
  \bibfield  {author} {\bibinfo {author} {\bibfnamefont {Will}\ \bibnamefont
  {Grathwohl}}, \bibinfo {author} {\bibfnamefont {Ricky T~Q}\ \bibnamefont
  {Chen}}, \bibinfo {author} {\bibfnamefont {Jesse}\ \bibnamefont
  {Bettencourt}}, \bibinfo {author} {\bibfnamefont {Ilya}\ \bibnamefont
  {Sutskever}}, \ and\ \bibinfo {author} {\bibfnamefont {David}\ \bibnamefont
  {Duvenaud}},\ }\bibfield  {title} {\enquote {\bibinfo {title} {{FFJORD:
  Free-form Continuous Dynamics for Scalable Reversible Generative Models}},}\
  }\href {http://arxiv.org/abs/1810.01367} {\  (\bibinfo {year} {2018})},\
  \Eprint {http://arxiv.org/abs/1810.01367} {arXiv:1810.01367} \BibitemShut
  {NoStop}%
\bibitem [{\citenamefont {Li}\ and\ \citenamefont {Wang}(2018)}]{Li2018z}%
  \BibitemOpen
  \bibfield  {author} {\bibinfo {author} {\bibfnamefont {Shuo-Hui}\
  \bibnamefont {Li}}\ and\ \bibinfo {author} {\bibfnamefont {Lei}\ \bibnamefont
  {Wang}},\ }\bibfield  {title} {\enquote {\bibinfo {title} {{Neural Network
  Renormalization Group}},}\ }\href {\doibase 10.1074/jbc.273.48.32230}
  {\bibfield  {journal} {\bibinfo  {journal} {Phys. Rev. Lett.}\ }\textbf
  {\bibinfo {volume} {121}},\ \bibinfo {pages} {260601} (\bibinfo {year}
  {2018})},\ \Eprint {http://arxiv.org/abs/1802.02840} {arXiv:1802.02840}
  \BibitemShut {NoStop}%
\bibitem [{\citenamefont {Hu}\ \emph {et~al.}(2019)\citenamefont {Hu},
  \citenamefont {Li}, \citenamefont {Wang},\ and\ \citenamefont
  {You}}]{Hu2019}%
  \BibitemOpen
  \bibfield  {author} {\bibinfo {author} {\bibfnamefont {Hong-Ye}\ \bibnamefont
  {Hu}}, \bibinfo {author} {\bibfnamefont {Shuo-Hui}\ \bibnamefont {Li}},
  \bibinfo {author} {\bibfnamefont {Lei}\ \bibnamefont {Wang}}, \ and\ \bibinfo
  {author} {\bibfnamefont {Yi-Zhuang}\ \bibnamefont {You}},\ }\bibfield
  {title} {\enquote {\bibinfo {title} {{Machine Learning Holographic Mapping by
  Neural Network Renormalization Group}},}\ }\href
  {http://arxiv.org/abs/1903.00804} {\  (\bibinfo {year} {2019})},\ \Eprint
  {http://arxiv.org/abs/1903.00804} {arXiv:1903.00804} \BibitemShut {NoStop}%
\bibitem [{\citenamefont {Song}\ \emph {et~al.}(2017)\citenamefont {Song},
  \citenamefont {Zhao},\ and\ \citenamefont {Ermon}}]{Song2017}%
  \BibitemOpen
  \bibfield  {author} {\bibinfo {author} {\bibfnamefont {Jiaming}\ \bibnamefont
  {Song}}, \bibinfo {author} {\bibfnamefont {Shengjia}\ \bibnamefont {Zhao}}, \
  and\ \bibinfo {author} {\bibfnamefont {Stefano}\ \bibnamefont {Ermon}},\
  }\bibfield  {title} {\enquote {\bibinfo {title} {{A-NICE-MC: Adversarial
  Training for MCMC}},}\ }\href {http://arxiv.org/abs/1706.07561} {\  (\bibinfo
  {year} {2017})},\ \Eprint {http://arxiv.org/abs/1706.07561}
  {arXiv:1706.07561} \BibitemShut {NoStop}%
\bibitem [{\citenamefont {Levy}\ \emph {et~al.}(2017)\citenamefont {Levy},
  \citenamefont {Hoffman},\ and\ \citenamefont {Sohl-Dickstein}}]{Levy2017}%
  \BibitemOpen
  \bibfield  {author} {\bibinfo {author} {\bibfnamefont {Daniel}\ \bibnamefont
  {Levy}}, \bibinfo {author} {\bibfnamefont {Matthew~D.}\ \bibnamefont
  {Hoffman}}, \ and\ \bibinfo {author} {\bibfnamefont {Jascha}\ \bibnamefont
  {Sohl-Dickstein}},\ }\bibfield  {title} {\enquote {\bibinfo {title}
  {{Generalizing Hamiltonian Monte Carlo with Neural Networks}},}\ }\href
  {http://arxiv.org/abs/1711.09268} {\  (\bibinfo {year} {2017})},\ \Eprint
  {http://arxiv.org/abs/1711.09268} {arXiv:1711.09268} \BibitemShut {NoStop}%
\bibitem [{\citenamefont {Albergo}\ \emph {et~al.}(2019)\citenamefont
  {Albergo}, \citenamefont {Kanwar},\ and\ \citenamefont
  {Shanahan}}]{Albergo2019}%
  \BibitemOpen
  \bibfield  {author} {\bibinfo {author} {\bibfnamefont {M.~S.}\ \bibnamefont
  {Albergo}}, \bibinfo {author} {\bibfnamefont {G.}~\bibnamefont {Kanwar}}, \
  and\ \bibinfo {author} {\bibfnamefont {P.~E.}\ \bibnamefont {Shanahan}},\
  }\bibfield  {title} {\enquote {\bibinfo {title} {{Flow-based generative
  models for Markov chain Monte Carlo in lattice field theory}},}\ }\href
  {\doibase 10.1103/PhysRevD.100.034515} {\bibfield  {journal} {\bibinfo
  {journal} {Phys. Rev. D}\ }\textbf {\bibinfo {volume} {100}},\ \bibinfo
  {pages} {34515} (\bibinfo {year} {2019})},\ \Eprint
  {http://arxiv.org/abs/1904.12072} {arXiv:1904.12072} \BibitemShut {NoStop}%
\bibitem [{\citenamefont {No{\'{e}}}\ \emph {et~al.}(2019)\citenamefont
  {No{\'{e}}}, \citenamefont {Olsson}, \citenamefont {K{\"{o}}hler},\ and\
  \citenamefont {Wu}}]{BoltzmannGenerator}%
  \BibitemOpen
  \bibfield  {author} {\bibinfo {author} {\bibfnamefont {Frank}\ \bibnamefont
  {No{\'{e}}}}, \bibinfo {author} {\bibfnamefont {Simon}\ \bibnamefont
  {Olsson}}, \bibinfo {author} {\bibfnamefont {Jonas}\ \bibnamefont
  {K{\"{o}}hler}}, \ and\ \bibinfo {author} {\bibfnamefont {Hao}\ \bibnamefont
  {Wu}},\ }\bibfield  {title} {\enquote {\bibinfo {title} {{Boltzmann
  Generators -- Sampling Equilibrium States of Many-Body Systems with Deep
  Learning}},}\ }\href {\doibase 10.1126/science.aaw1147} {\bibfield  {journal}
  {\bibinfo  {journal} {Science}\ }\textbf {\bibinfo {volume} {365}} (\bibinfo
  {year} {2019}),\ 10.1126/science.aaw1147},\ \Eprint
  {http://arxiv.org/abs/1812.01729} {arXiv:1812.01729} \BibitemShut {NoStop}%
\bibitem [{\citenamefont {Hartnett}\ and\ \citenamefont
  {Mohseni}(2020)}]{2001.00585}%
  \BibitemOpen
  \bibfield  {author} {\bibinfo {author} {\bibfnamefont {Gavin~S.}\
  \bibnamefont {Hartnett}}\ and\ \bibinfo {author} {\bibfnamefont {Masoud}\
  \bibnamefont {Mohseni}},\ }\bibfield  {title} {\enquote {\bibinfo {title}
  {Self-supervised learning of generative spin-glasses with normalizing
  flows},}\ }\href {http://arxiv.org/abs/2001.00585v1} {\bibfield  {journal}
  {\bibinfo  {journal} {CoRR}\ } (\bibinfo {year} {2020})},\ \Eprint
  {http://arxiv.org/abs/2001.00585} {arXiv:2001.00585 [cs.LG]} \BibitemShut
  {NoStop}%
\bibitem [{\citenamefont {E}(2017)}]{Weinan2017}%
  \BibitemOpen
  \bibfield  {author} {\bibinfo {author} {\bibfnamefont {Weinan}\ \bibnamefont
  {E}},\ }\bibfield  {title} {\enquote {\bibinfo {title} {{A Proposal on
  Machine Learning via Dynamical Systems}},}\ }\href {\doibase
  10.1007/s40304-017-0103-z} {\bibfield  {journal} {\bibinfo  {journal}
  {Commun. Math. Stat.}\ }\textbf {\bibinfo {volume} {5}},\ \bibinfo {pages}
  {1--11} (\bibinfo {year} {2017})}\BibitemShut {NoStop}%
\bibitem [{\citenamefont {Chen}\ \emph {et~al.}(2018)\citenamefont {Chen},
  \citenamefont {Rubanova}, \citenamefont {Bettencourt},\ and\ \citenamefont
  {Duvenaud}}]{NeuralODE}%
  \BibitemOpen
  \bibfield  {author} {\bibinfo {author} {\bibfnamefont {Tian~Qi}\ \bibnamefont
  {Chen}}, \bibinfo {author} {\bibfnamefont {Yulia}\ \bibnamefont {Rubanova}},
  \bibinfo {author} {\bibfnamefont {Jesse}\ \bibnamefont {Bettencourt}}, \ and\
  \bibinfo {author} {\bibfnamefont {David}\ \bibnamefont {Duvenaud}},\
  }\bibfield  {title} {\enquote {\bibinfo {title} {{Neural Ordinary
  Differential Equations}},}\ }\href {http://arxiv.org/abs/1806.07366} {\
  (\bibinfo {year} {2018})},\ \Eprint {http://arxiv.org/abs/1806.07366}
  {arXiv:1806.07366} \BibitemShut {NoStop}%
\bibitem [{\citenamefont {Zhang}\ \emph
  {et~al.}(2018{\natexlab{c}})\citenamefont {Zhang}, \citenamefont {E},\ and\
  \citenamefont {Wang}}]{Zhang2018r}%
  \BibitemOpen
  \bibfield  {author} {\bibinfo {author} {\bibfnamefont {Linfeng}\ \bibnamefont
  {Zhang}}, \bibinfo {author} {\bibfnamefont {Weinan}\ \bibnamefont {E}}, \
  and\ \bibinfo {author} {\bibfnamefont {Lei}\ \bibnamefont {Wang}},\
  }\bibfield  {title} {\enquote {\bibinfo {title} {{Monge-Amp$\backslash$`ere
  Flow for Generative Modeling}},}\ }\href {http://arxiv.org/abs/1809.10188} {\
   (\bibinfo {year} {2018}{\natexlab{c}})},\ \Eprint
  {http://arxiv.org/abs/1809.10188} {arXiv:1809.10188} \BibitemShut {NoStop}%
\bibitem [{\citenamefont {Sussman}\ and\ \citenamefont
  {Wisdom}(2014)}]{Sussman}%
  \BibitemOpen
  \bibfield  {author} {\bibinfo {author} {\bibfnamefont {GJ}~\bibnamefont
  {Sussman}}\ and\ \bibinfo {author} {\bibfnamefont {J}~\bibnamefont
  {Wisdom}},\ }\href {\doibase 10.1017/S0956796803234870} {\emph {\bibinfo
  {title} {{Structure and interpretation of classical mechanics}}}},\ \bibinfo
  {edition} {2nd}\ ed.\ (\bibinfo  {publisher} {The MIT Press},\ \bibinfo
  {year} {2014})\BibitemShut {NoStop}%
\bibitem [{\citenamefont {Baydin}\ \emph {et~al.}(2015)\citenamefont {Baydin},
  \citenamefont {Pearlmutter}, \citenamefont {Radul},\ and\ \citenamefont
  {Siskind}}]{Baydin2018}%
  \BibitemOpen
  \bibfield  {author} {\bibinfo {author} {\bibfnamefont {Atilim~G{\"u}nes}\
  \bibnamefont {Baydin}}, \bibinfo {author} {\bibfnamefont {Barak~A.}\
  \bibnamefont {Pearlmutter}}, \bibinfo {author} {\bibfnamefont
  {Alexey~Andreyevich}\ \bibnamefont {Radul}}, \ and\ \bibinfo {author}
  {\bibfnamefont {Jeffrey~Mark}\ \bibnamefont {Siskind}},\ }\bibfield  {title}
  {\enquote {\bibinfo {title} {{Automatic differentiation in machine learning:
  a survey}},}\ }\href {\doibase 10.1016/j.advwatres.2018.01.009} {\bibfield
  {journal} {\bibinfo  {journal} {J. Mach. Learn.}\ }\textbf {\bibinfo {volume}
  {18}},\ \bibinfo {pages} {1--43} (\bibinfo {year} {2015})},\ \Eprint
  {http://arxiv.org/abs/1502.05767} {arXiv:1502.05767} \BibitemShut {NoStop}%
\bibitem [{\citenamefont {van~den Oord}\ \emph {et~al.}(2017)\citenamefont
  {van~den Oord}, \citenamefont {Li}, \citenamefont {Babuschkin}, \citenamefont
  {Simonyan}, \citenamefont {Vinyals}, \citenamefont {Kavukcuoglu},
  \citenamefont {van~den Driessche}, \citenamefont {Lockhart}, \citenamefont
  {Cobo}, \citenamefont {Stimberg}, \citenamefont {Casagrande}, \citenamefont
  {Grewe}, \citenamefont {Noury}, \citenamefont {Dieleman}, \citenamefont
  {Elsen}, \citenamefont {Kalchbrenner}, \citenamefont {Zen}, \citenamefont
  {Graves}, \citenamefont {King}, \citenamefont {Walters}, \citenamefont
  {Belov},\ and\ \citenamefont {Hassabis}}]{Oord2017}%
  \BibitemOpen
  \bibfield  {author} {\bibinfo {author} {\bibfnamefont {Aaron}\ \bibnamefont
  {van~den Oord}}, \bibinfo {author} {\bibfnamefont {Yazhe}\ \bibnamefont
  {Li}}, \bibinfo {author} {\bibfnamefont {Igor}\ \bibnamefont {Babuschkin}},
  \bibinfo {author} {\bibfnamefont {Karen}\ \bibnamefont {Simonyan}}, \bibinfo
  {author} {\bibfnamefont {Oriol}\ \bibnamefont {Vinyals}}, \bibinfo {author}
  {\bibfnamefont {Koray}\ \bibnamefont {Kavukcuoglu}}, \bibinfo {author}
  {\bibfnamefont {George}\ \bibnamefont {van~den Driessche}}, \bibinfo {author}
  {\bibfnamefont {Edward}\ \bibnamefont {Lockhart}}, \bibinfo {author}
  {\bibfnamefont {Luis~C.}\ \bibnamefont {Cobo}}, \bibinfo {author}
  {\bibfnamefont {Florian}\ \bibnamefont {Stimberg}}, \bibinfo {author}
  {\bibfnamefont {Norman}\ \bibnamefont {Casagrande}}, \bibinfo {author}
  {\bibfnamefont {Dominik}\ \bibnamefont {Grewe}}, \bibinfo {author}
  {\bibfnamefont {Seb}\ \bibnamefont {Noury}}, \bibinfo {author} {\bibfnamefont
  {Sander}\ \bibnamefont {Dieleman}}, \bibinfo {author} {\bibfnamefont {Erich}\
  \bibnamefont {Elsen}}, \bibinfo {author} {\bibfnamefont {Nal}\ \bibnamefont
  {Kalchbrenner}}, \bibinfo {author} {\bibfnamefont {Heiga}\ \bibnamefont
  {Zen}}, \bibinfo {author} {\bibfnamefont {Alex}\ \bibnamefont {Graves}},
  \bibinfo {author} {\bibfnamefont {Helen}\ \bibnamefont {King}}, \bibinfo
  {author} {\bibfnamefont {Tom}\ \bibnamefont {Walters}}, \bibinfo {author}
  {\bibfnamefont {Dan}\ \bibnamefont {Belov}}, \ and\ \bibinfo {author}
  {\bibfnamefont {Demis}\ \bibnamefont {Hassabis}},\ }\bibfield  {title}
  {\enquote {\bibinfo {title} {{Parallel WaveNet: Fast High-Fidelity Speech
  Synthesis}},}\ }\href {http://arxiv.org/abs/1711.10433} {\  (\bibinfo {year}
  {2017})},\ \Eprint {http://arxiv.org/abs/1711.10433} {arXiv:1711.10433}
  \BibitemShut {NoStop}%
\bibitem [{\citenamefont {Mohamed}\ \emph {et~al.}(2019)\citenamefont
  {Mohamed}, \citenamefont {Rosca}, \citenamefont {Figurnov},\ and\
  \citenamefont {Mnih}}]{Mohamed2019}%
  \BibitemOpen
  \bibfield  {author} {\bibinfo {author} {\bibfnamefont {Shakir}\ \bibnamefont
  {Mohamed}}, \bibinfo {author} {\bibfnamefont {Mihaela}\ \bibnamefont
  {Rosca}}, \bibinfo {author} {\bibfnamefont {Michael}\ \bibnamefont
  {Figurnov}}, \ and\ \bibinfo {author} {\bibfnamefont {Andriy}\ \bibnamefont
  {Mnih}},\ }\bibfield  {title} {\enquote {\bibinfo {title} {{Monte Carlo
  Gradient Estimation in Machine Learning}},}\ }\href
  {http://arxiv.org/abs/1906.10652} {\  (\bibinfo {year} {2019})},\ \Eprint
  {http://arxiv.org/abs/1906.10652} {arXiv:1906.10652} \BibitemShut {NoStop}%
\bibitem [{\citenamefont {Wu}\ \emph {et~al.}(2018)\citenamefont {Wu},
  \citenamefont {Wang},\ and\ \citenamefont {Zhang}}]{Wu2018f}%
  \BibitemOpen
  \bibfield  {author} {\bibinfo {author} {\bibfnamefont {Dian}\ \bibnamefont
  {Wu}}, \bibinfo {author} {\bibfnamefont {Lei}\ \bibnamefont {Wang}}, \ and\
  \bibinfo {author} {\bibfnamefont {Pan}\ \bibnamefont {Zhang}},\ }\bibfield
  {title} {\enquote {\bibinfo {title} {{Solving Statistical Mechanics using
  Variational Autoregressive Networks}},}\ }\href {\doibase
  10.1103/PhysRevLett.122.080602} {\bibfield  {journal} {\bibinfo  {journal}
  {Phys. Rev. Lett.}\ }\textbf {\bibinfo {volume} {122}},\ \bibinfo {pages}
  {080602} (\bibinfo {year} {2018})},\ \Eprint
  {http://arxiv.org/abs/1809.10606} {arXiv:1809.10606} \BibitemShut {NoStop}%
\bibitem [{Git()}]{Github}%
  \BibitemOpen
  \href@noop {} {}\bibinfo {note} {See
  \href{https://github.com/li012589/neuralCT}{https://github.com/li012589/neuralCT}
  for code implementation in PyTorch}\BibitemShut {NoStop}%
\bibitem [{\citenamefont {Pearson}(1901)}]{Pearson:1901gs}%
  \BibitemOpen
  \bibfield  {author} {\bibinfo {author} {\bibfnamefont {Karl}\ \bibnamefont
  {Pearson}},\ }\bibfield  {title} {{\selectlanguage {English}\enquote
  {\bibinfo {title} {{LIII. On lines and planes of closest fit to systems of
  points in space}},}\ }}\href {\doibase 10.1080/14786440109462720} {\bibfield
  {journal} {\bibinfo  {journal} {Philosophical Magazine}\ }\textbf {\bibinfo
  {volume} {2}},\ \bibinfo {pages} {559--572} (\bibinfo {year}
  {1901})}\BibitemShut {NoStop}%
\bibitem [{\citenamefont {N{\"{u}}ske}\ \emph {et~al.}(2017)\citenamefont
  {N{\"{u}}ske}, \citenamefont {Wu}, \citenamefont {Prinz}, \citenamefont
  {Wehmeyer}, \citenamefont {Clementi},\ and\ \citenamefont
  {No{\'{e}}}}]{Nuske2017}%
  \BibitemOpen
  \bibfield  {author} {\bibinfo {author} {\bibfnamefont {Feliks}\ \bibnamefont
  {N{\"{u}}ske}}, \bibinfo {author} {\bibfnamefont {Hao}\ \bibnamefont {Wu}},
  \bibinfo {author} {\bibfnamefont {Jan~Hendrik}\ \bibnamefont {Prinz}},
  \bibinfo {author} {\bibfnamefont {Christoph}\ \bibnamefont {Wehmeyer}},
  \bibinfo {author} {\bibfnamefont {Cecilia}\ \bibnamefont {Clementi}}, \ and\
  \bibinfo {author} {\bibfnamefont {Frank}\ \bibnamefont {No{\'{e}}}},\
  }\bibfield  {title} {\enquote {\bibinfo {title} {{Markov state models from
  short non-equilibrium simulations - Analysis and correction of estimation
  bias}},}\ }\href {https://doi.org/10.1063/1.4976518} {\bibfield  {journal}
  {\bibinfo  {journal} {J. Chem. Phys.}\ }\textbf {\bibinfo {volume} {146}},\
  \bibinfo {pages} {094104} (\bibinfo {year} {2017})}\BibitemShut {NoStop}%
\bibitem [{mds()}]{mdshare}%
  \BibitemOpen
  \href@noop {} {}\bibinfo {note}
  {\href{https://markovmodel.github.io/mdshare/ALA2/\#alanine-dipeptide}{https://markovmodel.github.io/mdshare/ALA2/\#alanine-dipeptide}}\BibitemShut
  {NoStop}%
\bibitem [{\citenamefont {Kingma}\ and\ \citenamefont
  {Ba}(2015)}]{kingma2015adam}%
  \BibitemOpen
  \bibfield  {author} {\bibinfo {author} {\bibfnamefont {Diederik}\
  \bibnamefont {Kingma}}\ and\ \bibinfo {author} {\bibfnamefont {Jimmy}\
  \bibnamefont {Ba}},\ }\bibfield  {title} {\enquote {\bibinfo {title} {Adam: a
  method for stochastic optimization},}\ }in\ \href@noop {} {\emph {\bibinfo
  {booktitle} {Proceedings of the International Conference on Learning
  Representations (ICLR)}}}\ (\bibinfo {year} {2015})\BibitemShut {NoStop}%
\bibitem [{\citenamefont {Kraskov}\ \emph {et~al.}(2004)\citenamefont
  {Kraskov}, \citenamefont {St{\"{o}}gbauer},\ and\ \citenamefont
  {Grassberger}}]{Kraskov2004}%
  \BibitemOpen
  \bibfield  {author} {\bibinfo {author} {\bibfnamefont {Alexander}\
  \bibnamefont {Kraskov}}, \bibinfo {author} {\bibfnamefont {Harald}\
  \bibnamefont {St{\"{o}}gbauer}}, \ and\ \bibinfo {author} {\bibfnamefont
  {Peter}\ \bibnamefont {Grassberger}},\ }\bibfield  {title} {\enquote
  {\bibinfo {title} {{Estimating mutual information}},}\ }\href {\doibase
  10.1103/PhysRevE.69.066138} {\bibfield  {journal} {\bibinfo  {journal} {Phys.
  Rev. E}\ }\textbf {\bibinfo {volume} {69}},\ \bibinfo {pages} {066138}
  (\bibinfo {year} {2004})}\BibitemShut {NoStop}%
\bibitem [{\citenamefont {White}(2016)}]{slerp}%
  \BibitemOpen
  \bibfield  {author} {\bibinfo {author} {\bibfnamefont {Tom}\ \bibnamefont
  {White}},\ }\bibfield  {title} {\enquote {\bibinfo {title} {Sampling
  generative networks},}\ }\href@noop {} {\  (\bibinfo {year} {2016})},\
  \Eprint {http://arxiv.org/abs/1609.04468} {arXiv:1609.04468 [cs.NE]}
  \BibitemShut {NoStop}%
\bibitem [{\citenamefont {Huang}\ and\ \citenamefont
  {Wang}(2017)}]{Huang2017a}%
  \BibitemOpen
  \bibfield  {author} {\bibinfo {author} {\bibfnamefont {Li}~\bibnamefont
  {Huang}}\ and\ \bibinfo {author} {\bibfnamefont {Lei}\ \bibnamefont {Wang}},\
  }\bibfield  {title} {\enquote {\bibinfo {title} {{Accelerated Monte Carlo
  simulations with restricted Boltzmann machines}},}\ }\href {\doibase
  10.1103/PhysRevB.95.035105} {\bibfield  {journal} {\bibinfo  {journal} {Phys.
  Rev. B}\ }\textbf {\bibinfo {volume} {95}},\ \bibinfo {pages} {035105}
  (\bibinfo {year} {2017})},\ \Eprint {http://arxiv.org/abs/1610.02746}
  {arXiv:1610.02746} \BibitemShut {NoStop}%
\bibitem [{\citenamefont {Liu}\ \emph {et~al.}(2017)\citenamefont {Liu},
  \citenamefont {Qi}, \citenamefont {Meng},\ and\ \citenamefont
  {Fu}}]{Liu2017f}%
  \BibitemOpen
  \bibfield  {author} {\bibinfo {author} {\bibfnamefont {Junwei}\ \bibnamefont
  {Liu}}, \bibinfo {author} {\bibfnamefont {Yang}\ \bibnamefont {Qi}}, \bibinfo
  {author} {\bibfnamefont {Zi~Yang}\ \bibnamefont {Meng}}, \ and\ \bibinfo
  {author} {\bibfnamefont {Liang}\ \bibnamefont {Fu}},\ }\bibfield  {title}
  {\enquote {\bibinfo {title} {{Self-learning Monte Carlo method}},}\ }\href
  {\doibase 10.1103/PhysRevB.95.041101} {\bibfield  {journal} {\bibinfo
  {journal} {Phys. Rev. B}\ }\textbf {\bibinfo {volume} {95}},\ \bibinfo
  {pages} {041101(R)} (\bibinfo {year} {2017})},\ \Eprint
  {http://arxiv.org/abs/1610.03137} {arXiv:1610.03137} \BibitemShut {NoStop}%
\bibitem [{Note1()}]{Note1}%
  \BibitemOpen
  \bibinfo {note} {See Ref.~\cite {Zhang2018r} for the preprocessing steps to
  map MNIST dataset to continuous variables.}\BibitemShut {Stop}%
\bibitem [{\citenamefont {Gregor}\ \emph
  {et~al.}(2016{\natexlab{b}})\citenamefont {Gregor}, \citenamefont {Besse},
  \citenamefont {Rezende}, \citenamefont {Danihelka},\ and\ \citenamefont
  {Wierstra}}]{Gregor2016}%
  \BibitemOpen
  \bibfield  {author} {\bibinfo {author} {\bibfnamefont {Karol}\ \bibnamefont
  {Gregor}}, \bibinfo {author} {\bibfnamefont {Frederic}\ \bibnamefont
  {Besse}}, \bibinfo {author} {\bibfnamefont {Danilo~Jimenez}\ \bibnamefont
  {Rezende}}, \bibinfo {author} {\bibfnamefont {Ivo}\ \bibnamefont
  {Danihelka}}, \ and\ \bibinfo {author} {\bibfnamefont {Daan}\ \bibnamefont
  {Wierstra}},\ }\bibfield  {title} {\enquote {\bibinfo {title} {{Towards
  Conceptual Compression}},}\ }\href {http://arxiv.org/abs/1604.08772} {\
  (\bibinfo {year} {2016}{\natexlab{b}})},\ \Eprint
  {http://arxiv.org/abs/1604.08772} {arXiv:1604.08772} \BibitemShut {NoStop}%
\bibitem [{\citenamefont {Laio}\ and\ \citenamefont
  {Parrinello}(2002)}]{Laio2002}%
  \BibitemOpen
  \bibfield  {author} {\bibinfo {author} {\bibfnamefont {Alessandro}\
  \bibnamefont {Laio}}\ and\ \bibinfo {author} {\bibfnamefont {Michele}\
  \bibnamefont {Parrinello}},\ }\bibfield  {title} {\enquote {\bibinfo {title}
  {{Escaping free-energy minima}},}\ }\href
  {https://www.pnas.org/content/99/20/12562} {\bibfield  {journal} {\bibinfo
  {journal} {PNAS}\ }\textbf {\bibinfo {volume} {99}},\ \bibinfo {pages}
  {12562} (\bibinfo {year} {2002})}\BibitemShut {NoStop}%
\bibitem [{\citenamefont {Valsson}\ and\ \citenamefont
  {Parrinello}(2014)}]{Valsson2014}%
  \BibitemOpen
  \bibfield  {author} {\bibinfo {author} {\bibfnamefont {Omar}\ \bibnamefont
  {Valsson}}\ and\ \bibinfo {author} {\bibfnamefont {Michele}\ \bibnamefont
  {Parrinello}},\ }\bibfield  {title} {\enquote {\bibinfo {title} {{Variational
  approach to enhanced sampling and free energy calculations}},}\ }\href
  {\doibase 10.1103/PhysRevLett.113.090601} {\bibfield  {journal} {\bibinfo
  {journal} {Phys. Rev. Lett.}\ }\textbf {\bibinfo {volume} {113}},\ \bibinfo
  {pages} {090601} (\bibinfo {year} {2014})}\BibitemShut {NoStop}%
\bibitem [{\citenamefont {Zhang}\ \emph
  {et~al.}(2018{\natexlab{d}})\citenamefont {Zhang}, \citenamefont {Wang},\
  and\ \citenamefont {E}}]{Zhang2018RiD}%
  \BibitemOpen
  \bibfield  {author} {\bibinfo {author} {\bibfnamefont {Linfeng}\ \bibnamefont
  {Zhang}}, \bibinfo {author} {\bibfnamefont {Han}\ \bibnamefont {Wang}}, \
  and\ \bibinfo {author} {\bibfnamefont {Weinan}\ \bibnamefont {E}},\
  }\bibfield  {title} {\enquote {\bibinfo {title} {Reinforced dynamics for
  enhanced sampling in large atomic and molecular systems},}\ }\href {\doibase
  10.1063/1.5019675} {\bibfield  {journal} {\bibinfo  {journal} {The Journal of
  Chemical Physics}\ }\textbf {\bibinfo {volume} {148}},\ \bibinfo {pages}
  {124113} (\bibinfo {year} {2018}{\natexlab{d}})},\ \Eprint
  {http://arxiv.org/abs/https://doi.org/10.1063/1.5019675}
  {https://doi.org/10.1063/1.5019675} \BibitemShut {NoStop}%
\bibitem [{\citenamefont {Zhu}\ \emph {et~al.}(2002)\citenamefont {Zhu},
  \citenamefont {Tuckerman}, \citenamefont {Samuelson},\ and\ \citenamefont
  {Martyna}}]{Zhu2002}%
  \BibitemOpen
  \bibfield  {author} {\bibinfo {author} {\bibfnamefont {Zhongwei}\
  \bibnamefont {Zhu}}, \bibinfo {author} {\bibfnamefont {Mark~E.}\ \bibnamefont
  {Tuckerman}}, \bibinfo {author} {\bibfnamefont {Shane~O.}\ \bibnamefont
  {Samuelson}}, \ and\ \bibinfo {author} {\bibfnamefont {Glenn~J.}\
  \bibnamefont {Martyna}},\ }\bibfield  {title} {\enquote {\bibinfo {title}
  {{Using Novel Variable Transformations to Enhance Conformational Sampling in
  Molecular Dynamics}},}\ }\href {\doibase 10.1103/physrevlett.88.100201}
  {\bibfield  {journal} {\bibinfo  {journal} {Phys. Rev. Lett.}\ }\textbf
  {\bibinfo {volume} {88}},\ \bibinfo {pages} {100201} (\bibinfo {year}
  {2002})}\BibitemShut {NoStop}%
\bibitem [{\citenamefont {Salimans}\ \emph {et~al.}(2014)\citenamefont
  {Salimans}, \citenamefont {Kingma},\ and\ \citenamefont
  {Welling}}]{Salimans2014}%
  \BibitemOpen
  \bibfield  {author} {\bibinfo {author} {\bibfnamefont {Tim}\ \bibnamefont
  {Salimans}}, \bibinfo {author} {\bibfnamefont {Diederik~P.}\ \bibnamefont
  {Kingma}}, \ and\ \bibinfo {author} {\bibfnamefont {Max}\ \bibnamefont
  {Welling}},\ }\bibfield  {title} {\enquote {\bibinfo {title} {{Markov Chain
  Monte Carlo and Variational Inference: Bridging the Gap}},}\ }\href
  {http://arxiv.org/abs/1410.6460} {\  (\bibinfo {year} {2014})},\ \Eprint
  {http://arxiv.org/abs/1410.6460} {arXiv:1410.6460} \BibitemShut {NoStop}%
\bibitem [{\citenamefont {Chen}\ \emph {et~al.}(2016)\citenamefont {Chen},
  \citenamefont {Duan}, \citenamefont {Houthooft}, \citenamefont {Schulman},
  \citenamefont {Sutskever},\ and\ \citenamefont {Abbeel}}]{Chen2016b}%
  \BibitemOpen
  \bibfield  {author} {\bibinfo {author} {\bibfnamefont {Xi}~\bibnamefont
  {Chen}}, \bibinfo {author} {\bibfnamefont {Yan}\ \bibnamefont {Duan}},
  \bibinfo {author} {\bibfnamefont {Rein}\ \bibnamefont {Houthooft}}, \bibinfo
  {author} {\bibfnamefont {John}\ \bibnamefont {Schulman}}, \bibinfo {author}
  {\bibfnamefont {Ilya}\ \bibnamefont {Sutskever}}, \ and\ \bibinfo {author}
  {\bibfnamefont {Pieter}\ \bibnamefont {Abbeel}},\ }\bibfield  {title}
  {\enquote {\bibinfo {title} {{InfoGAN: Interpretable Representation Learning
  by Information Maximizing Generative Adversarial Nets}},}\ }\href
  {http://arxiv.org/abs/1606.03657} {\  (\bibinfo {year} {2016})},\ \Eprint
  {http://arxiv.org/abs/1606.03657} {arXiv:1606.03657} \BibitemShut {NoStop}%
\bibitem [{\citenamefont {Higgins}\ \emph {et~al.}(2017)\citenamefont
  {Higgins}, \citenamefont {Matthey}, \citenamefont {Pal}, \citenamefont
  {Burgess}, \citenamefont {Glorot}, \citenamefont {Botvinick}, \citenamefont
  {Mohamed},\ and\ \citenamefont {Lerchner}}]{higgins2017beta}%
  \BibitemOpen
  \bibfield  {author} {\bibinfo {author} {\bibfnamefont {Irina}\ \bibnamefont
  {Higgins}}, \bibinfo {author} {\bibfnamefont {Loic}\ \bibnamefont {Matthey}},
  \bibinfo {author} {\bibfnamefont {Arka}\ \bibnamefont {Pal}}, \bibinfo
  {author} {\bibfnamefont {Christopher}\ \bibnamefont {Burgess}}, \bibinfo
  {author} {\bibfnamefont {Xavier}\ \bibnamefont {Glorot}}, \bibinfo {author}
  {\bibfnamefont {Matthew}\ \bibnamefont {Botvinick}}, \bibinfo {author}
  {\bibfnamefont {Shakir}\ \bibnamefont {Mohamed}}, \ and\ \bibinfo {author}
  {\bibfnamefont {Alexander}\ \bibnamefont {Lerchner}},\ }\bibfield  {title}
  {\enquote {\bibinfo {title} {beta-vae: Learning basic visual concepts with a
  constrained variational framework},}\ }in\ \href@noop {} {\emph {\bibinfo
  {booktitle} {International Conference on Learning Representations}}},\
  Vol.~\bibinfo {volume} {3}\ (\bibinfo {year} {2017})\BibitemShut {NoStop}%
\bibitem [{\citenamefont {Karras}\ \emph {et~al.}(2018)\citenamefont {Karras},
  \citenamefont {Laine},\ and\ \citenamefont {Aila}}]{Karras2018}%
  \BibitemOpen
  \bibfield  {author} {\bibinfo {author} {\bibfnamefont {Tero}\ \bibnamefont
  {Karras}}, \bibinfo {author} {\bibfnamefont {Samuli}\ \bibnamefont {Laine}},
  \ and\ \bibinfo {author} {\bibfnamefont {Timo}\ \bibnamefont {Aila}},\
  }\bibfield  {title} {\enquote {\bibinfo {title} {{A Style-Based Generator
  Architecture for Generative Adversarial Networks}},}\ }\href
  {http://arxiv.org/abs/1812.04948} {\  (\bibinfo {year} {2018})},\ \Eprint
  {http://arxiv.org/abs/1812.04948} {arXiv:1812.04948} \BibitemShut {NoStop}%
\bibitem [{\citenamefont {Das}\ \emph {et~al.}(2019)\citenamefont {Das},
  \citenamefont {Abbeel},\ and\ \citenamefont {Spanos}}]{Das2019}%
  \BibitemOpen
  \bibfield  {author} {\bibinfo {author} {\bibfnamefont {Hari~Prasanna}\
  \bibnamefont {Das}}, \bibinfo {author} {\bibfnamefont {Pieter}\ \bibnamefont
  {Abbeel}}, \ and\ \bibinfo {author} {\bibfnamefont {Costas~J}\ \bibnamefont
  {Spanos}},\ }\bibfield  {title} {\enquote {\bibinfo {title} {{Dimensionality
  Reduction Flows}},}\ }\href {http://arxiv.org/abs/1908.01686} {\  (\bibinfo
  {year} {2019})},\ \Eprint {http://arxiv.org/abs/1908.01686}
  {arXiv:1908.01686} \BibitemShut {NoStop}%
\bibitem [{\citenamefont {K{\"{o}}hler}\ \emph {et~al.}(2019)\citenamefont
  {K{\"{o}}hler}, \citenamefont {Klein},\ and\ \citenamefont
  {No{\'{e}}}}]{Kohlerb}%
  \BibitemOpen
  \bibfield  {author} {\bibinfo {author} {\bibfnamefont {Jonas}\ \bibnamefont
  {K{\"{o}}hler}}, \bibinfo {author} {\bibfnamefont {Leon}\ \bibnamefont
  {Klein}}, \ and\ \bibinfo {author} {\bibfnamefont {Frank}\ \bibnamefont
  {No{\'{e}}}},\ }\bibfield  {title} {\enquote {\bibinfo {title} {{Equivariant
  Flows: sampling configurations for multi-body systems with symmetric
  energies}},}\ }\href {http://arxiv.org/abs/1910.00753} {\  (\bibinfo {year}
  {2019})},\ \Eprint {http://arxiv.org/abs/1910.00753} {arXiv:1910.00753}
  \BibitemShut {NoStop}%
\bibitem [{\citenamefont {Dumas}(2014)}]{dumas2014kam}%
  \BibitemOpen
  \bibfield  {author} {\bibinfo {author} {\bibfnamefont {H~Scott}\ \bibnamefont
  {Dumas}},\ }\href@noop {} {\emph {\bibinfo {title} {The KAM Story: A Friendly
  Introduction to the Content, History, and Significance of Classical
  Kolmogorov-Arnold-Moser Theory}}}\ (\bibinfo  {publisher} {World Scientific
  Publishing Company},\ \bibinfo {year} {2014})\BibitemShut {NoStop}%
\bibitem [{\citenamefont {Acebr\'on}\ \emph {et~al.}(2005)\citenamefont
  {Acebr\'on}, \citenamefont {Bonilla}, \citenamefont {P\'erez~Vicente},
  \citenamefont {Ritort},\ and\ \citenamefont {Spigler}}]{RevModPhys.77.137}%
  \BibitemOpen
  \bibfield  {author} {\bibinfo {author} {\bibfnamefont {Juan~A.}\ \bibnamefont
  {Acebr\'on}}, \bibinfo {author} {\bibfnamefont {L.~L.}\ \bibnamefont
  {Bonilla}}, \bibinfo {author} {\bibfnamefont {Conrad~J.}\ \bibnamefont
  {P\'erez~Vicente}}, \bibinfo {author} {\bibfnamefont {F\'elix}\ \bibnamefont
  {Ritort}}, \ and\ \bibinfo {author} {\bibfnamefont {Renato}\ \bibnamefont
  {Spigler}},\ }\bibfield  {title} {\enquote {\bibinfo {title} {The kuramoto
  model: A simple paradigm for synchronization phenomena},}\ }\href {\doibase
  10.1103/RevModPhys.77.137} {\bibfield  {journal} {\bibinfo  {journal} {Rev.
  Mod. Phys.}\ }\textbf {\bibinfo {volume} {77}},\ \bibinfo {pages} {137--185}
  (\bibinfo {year} {2005})}\BibitemShut {NoStop}%
\bibitem [{\citenamefont {Hall}(2015)}]{Hall2015}%
  \BibitemOpen
  \bibfield  {author} {\bibinfo {author} {\bibfnamefont {Brian}\ \bibnamefont
  {Hall}},\ }\href {\doibase 10.1007/978-0-8176-4493-2_1} {\emph {\bibinfo
  {title} {{Lie Groups, Lie Algebras and Representations}}}},\ \bibinfo
  {edition} {2nd}\ ed.\ (\bibinfo  {publisher} {Springer},\ \bibinfo {year}
  {2015})\BibitemShut {NoStop}%
\bibitem [{\citenamefont {Lezcano-Casado}(2019)}]{Lezcano-Casado2019}%
  \BibitemOpen
  \bibfield  {author} {\bibinfo {author} {\bibfnamefont {Mario}\ \bibnamefont
  {Lezcano-Casado}},\ }\bibfield  {title} {\enquote {\bibinfo {title}
  {{Trivializations for Gradient-Based Optimization on Manifolds}},}\ }\href
  {http://arxiv.org/abs/1909.09501} {\  (\bibinfo {year} {2019})},\ \Eprint
  {http://arxiv.org/abs/1909.09501} {arXiv:1909.09501} \BibitemShut {NoStop}%
\end{thebibliography}%

\clearpage 
\appendix 

\section{Symplectic Flows} \label{sec:flows}
We list several other forms of neural symplectic transformation and discuss their relation to known constructions in the literature. 

\subsection{Linear symplectic transformation}
The simplest canonical transformation is a linear transformation to the input variables. We parameterize the linear symplectic transformation using the exponential map of its Lie algebra~\cite{Hall2015},   
\begin{equation} 
\vlatent =  \vphysical\, e^{Y} \quad \mathrm{with} \quad Y={\left(\begin{array}{cc}A & B \\ C &-A^T \end{array}\right)}, 
\label{eq:liealg} 
\end{equation}
where $B, C$ are real symmetric matrices and $A$ is an arbitrary real matrix. One can implement \Eq{eq:liealg} via efficient vector-matrix exponential multiplication. Since the symplectic group is connected~\cite{Hall2015}, the exponential map covers all linear symplectic transformations. Moreover, one can obtain the reverse of the transformation by acting $ e^{-Y}$ instead. Accurate and efficient differentiation through the matrix exponential is discussed in Ref.~\cite{Lezcano-Casado2019}. 

In the special case of $B=C=0$ and $A$ is a skew-symmetric matrix, i.e. $A=-A^T$, the linear symplectic transformation reduces to the orthogonal transformation of both momenta and coordinates, which corresponds to the normal mode transformation. 

\subsection{Continuous symplectic flow}
In general, one can parameterize the canonical transformation using a scalar generating function $G(\vlambda)$. Integrating the ordinary differential equation (ODE) 
%\begin{eqnarray}
%\frac{d p}{d t} & =& \frac{\partial G}{\partial q} \\
%\frac{d q}{d t} & =& -\frac{\partial G}{\partial p} 
%\end{eqnarray}
\begin{equation}
\dot{\vlambda} = \nabla_{\vlambda} G(\vlambda) J
\label{eq:symplecticflow}
\end{equation}
from time $0$ to $\tau$, 
one can transform the original variables from $\vlambda(t=0)=\vphysical$ to  $\vlambda(t=\tau)=\vlatent$. As a consequence, the Hamiltonian evolution is a special form of symplectic flow in the phase space with the generating function being the Hamiltonian~\cite{Arnold1989}. 
%Both the Hamiltonian evolution and the canonical transformation are a symplectic flow in the phase space. 
%This class of symplectic transformation employs the fact that time evolution and a given Hamiltonian is also a symplectic transformation.
%We discuss practical implementations of the neural canonical transformation in Sec.~\ref{sec:implementation}. 
%A conniventi case is the the Storm-Verlet, which originates from an symplectic integrator for a separable Hamiltonain $G(p, q) = T(p)+V(q)$
%\begin{eqnarray}
%p_mid = p - \frac{1}{2} \frac{\partial V(q)}{\partial q} \\
%Q = q+ \frac{\partial T(p_mid)}{\partial p} \\
%P = p -\frac{1}{2} \frac{\partial V}{\partial q}\Bigr|_Q
%\end{eqnarray}
%This transformation is easily reversible. 

Equation~(\ref{eq:symplecticflow}) corresponds to the infinitesimal canonical transformation~\cite{Arnold1989}, which covers a broad family of symplectic transformations discussed so far. For example, if the generating function is a linear function of the momenta, we will arrive at the neural point transformation Eqs.~(\ref{eq:coord}, \ref{eq:momentum}) introduced in the main texts. While if the generating function is a quadratic function of $\vlambda$ we obtain the linear symplectic transformation \Eq{eq:liealg}. 
 
The continuous symplectic flow falls into the framework of \MA flow in the  optimal transport theory~\cite{Zhang2018r}, where the transportation is induced by a gradient flow under a scalar potential function. %In contrast to the more general \MA flow discussed in~\cite{Zhang2018r}, here one has an irrotational and incompressible flow in the phase space. 
The symplectic structure in \Eq{eq:symplecticflow} simplifies the computation %since one has $d  \rho(\vlambda, t)/dt = 0$  
due to the volume-preserving property. In practice, the continuous transformation \Eq{eq:symplecticflow} can be implemented via the neural ODE~\cite{NeuralODE}. Since the symplectic symmetry is crucial for the canonical transformation, it is crucial to employ symplectic integrators~\cite{Feng2011} in the neural ODE implementation. In particular, if one  employs a symplectic leap-frog discretization of the \Eq{eq:symplecticflow} for a separable generating function, one will arrive at the transformations discussed in ~\cite{Levy2017, Bondesan2019}. %While a general generating function does not even need to be separable, which often requires implicit integration scheme with the symplectic condition.

\section{Conceptual compression of the MNIST dataset} \label{app:compress}
\begin{figure}[h!]
    \begin{center}
      \includegraphics[width=\columnwidth, trim={0 0cm 0cm 0cm}, clip]{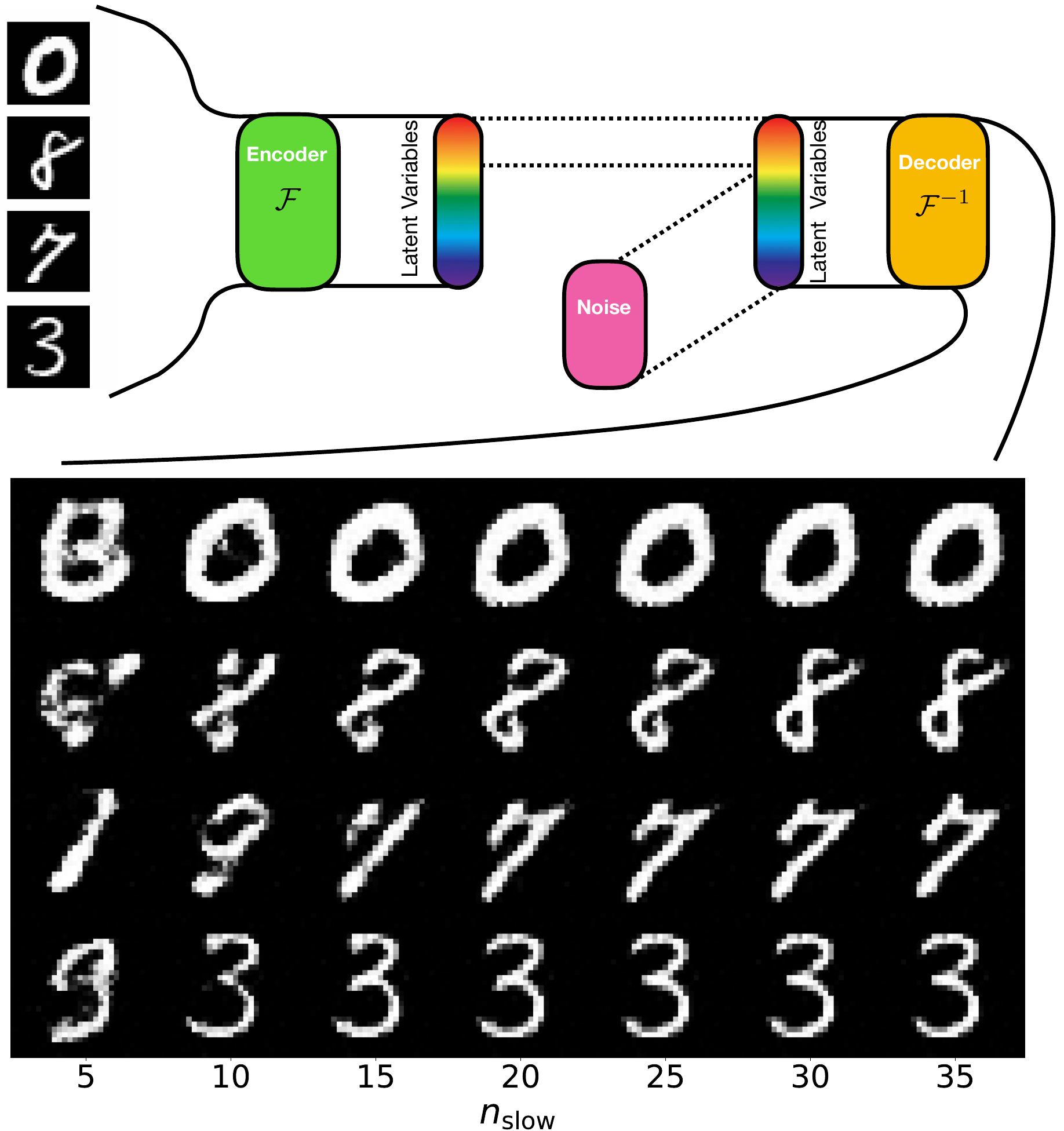} 
    \end{center}
    \caption{Conceptual compression using the learned collective variables. One performs the coordinate transformation \Eq{eq:coord} to the input data, and restores the data based on $n_\mathrm{slow}$ slowest modes. The remaining fast modes are thrown away and resampled from the prior distribution.}\label{fig:compress_concept}
\end{figure}

The extraction of the salient features as the slow modes is useful for compressions. For example, conceptual compression is a lossy compression scheme that aims at capturing the global information of the input data~\cite{Gregor2016, Dinh2016a}. 
The conventional approaches make use of the VAEs or the neural networks with a hierarchical structure. 
However,  we  perform the compression based on the learned frequencies since the symplectic network naturally separates fast and slow degrees of freedom of the dataset. The top of Fig.~\ref{fig:compress_concept} shows the setup of conceptual compression with learned neural canonical transformation. 
First, we use the learned nonlinear coordinate transformation \Eq{eq:coord} to map the data to the latent space. 
Then, we only pass a few slow modes to a decoder network. 
The decoder restores the image from the latent space by running the inverse transformation as the encoder network. 
To make up the missing information, we simply sample the high-frequency modes from the prior distribution and feed them into the decoder. 
The bottom of Figure~\ref{fig:compress_concept} shows the results of the conceptual compression, that from left to right we keep $5,10,15,20,25,30,35$ of the slowest collective variables. 
The conceptual compression experiments show that the symplectic transformation captures the global information in the slow modes in the latent space since one can restore the image with only a small number of the slowest variables.

\end{document}